\documentclass[sigconf]{acmart}
\AtBeginDocument{%
  \providecommand\BibTeX{{%
    \normalfont B\kern-0.5em{\scshape i\kern-0.25em b}\kern-0.8em\TeX}}}
\usepackage{graphicx,tabularx,subcaption}
\usepackage[capitalise,nameinlink,compress]{cleveref}
\usepackage{xcolor,mathtools}
\usepackage[dutch,english]{babel}
\usepackage{acmart-taps}

\graphicspath{{images/}}

\newcommand{\insignif}[1]{\textcolor{lightgray}{#1}}
\newcommand{\dutch}[1]{\foreignlanguage{dutch}{#1}}
\aptLtoXcmd{
\def\citeg#1{\cite{#1}}
}{
\def\citeg#1{\cite[e.g.,][]{#1}}
}
\renewcommand{\quote}[1]{\textit{``#1''}}
\DeclareMathOperator{\elo}{Elo}

\def\anon#1{#1}
\usepackage{acmart-taps}
\copyrightyear{2023} 
\acmYear{2023} 
\setcopyright{acmlicensed}\acmConference[IUI '23]{28th International Conference on Intelligent User Interfaces}{March 27--31, 2023}{Sydney, NSW, Australia}
\acmBooktitle{28th International Conference on Intelligent User Interfaces (IUI '23), March 27--31, 2023, Sydney, NSW, Australia}
\acmPrice{15.00}
\acmDOI{10.1145/3581641.3584046}
\acmISBN{979-8-4007-0106-1/23/03}

\begin{document}

\title[Steering Recommendations and Visualising Its Impact]{Steering Recommendations and Visualising Its Impact: Effects on Adolescents' Trust in E-Learning Platforms}

\author{Jeroen Ooge}
\orcid{0000-0001-9820-7656}
\affiliation{%
  \institution{KU Leuven}
  \department{Department of Computer Science}
  \streetaddress{Celestijnenlaan 200A}
  \city{Leuven}
  \postcode{3001}
  \country{Belgium}
}
\email{jeroen.ooge@kuleuven.be}
\author{Leen Dereu}
\orcid{0000-0002-8330-0935}
\affiliation{%
  \institution{KU Leuven}
  \department{Department of Computer Science}
  \streetaddress{Celestijnenlaan 200A}
  \city{Leuven}
  \postcode{3001}
  \country{Belgium}
}
\email{leen.dereu@student.kuleuven.be}
\author{Katrien Verbert}
\orcid{0000-0001-6699-7710}
\affiliation{%
  \institution{KU Leuven}
  \department{Department of Computer Science}
  \streetaddress{Celestijnenlaan 200A}
  \city{Leuven}
  \postcode{3001}
  \country{Belgium}
}
\email{katrien.verbert@kuleuven.be}


\begin{abstract}
Researchers have widely acknowledged the potential of control mechanisms with which end-users of recommender systems can better tailor recommendations. However, few e-learning environments so far incorporate such mechanisms, for example for steering recommended exercises. In addition, studies with adolescents in this context are rare. To address these limitations, we designed a control mechanism and a visualisation of the control's impact through an iterative design process with adolescents and teachers. Then, we investigated how these functionalities affect adolescents' trust in an e-learning platform that recommends maths exercises. A randomised controlled experiment with 76~middle school and high school adolescents showed that visualising the impact of exercised control significantly increases trust. Furthermore, having control over their mastery level seemed to inspire adolescents to reasonably challenge themselves and reflect upon the underlying recommendation algorithm. Finally, a significant increase in perceived transparency suggested that visualising steering actions can indirectly explain why recommendations are suitable, which opens interesting research tracks for the broader field of explainable AI.
\end{abstract}




\maketitle

\section{Introduction}
Recommender systems have long been actively studied to help reduce information overload in contexts where people are searching for relevant content. To better anticipate people's changing preferences and needs, researchers have increasingly acknowledged the importance of control mechanisms with which people can actively steer recommendations~\cite{jannach2017user}. Studies have shown that being able to control recommendations can increase satisfaction with, perceived understanding of, and trust in a recommender system, which can in turn increase acceptance of recommendations~\cite{knijnenburg2012inspectability}. At the same time, too much control can overwhelm people and incur high cognitive loads~\cite{andjelkovic2016moodplay,bollen2010understanding}.

However, most research on controlling recommender systems is limited because of two reasons. First, studied target audiences typically consist of adults, whereas in practice younger audiences such as adolescents (ages 12--19~\cite{fitton2013challenge}) are just as much, if not more, exposed to recommendation algorithms. Second, recommender systems are most often studied within contexts such as multimedia, e-commerce, and other services, and it is unclear whether findings therein always transfer to other application domains. In a high-stakes domain such as education, for example, it is crucial to properly understand the effects of control mechanisms, especially now that e-learning platforms are increasingly recommending learning content to personalise learning. Thus, it is important to design control mechanisms fit for an educational context; reflect on how much control students, teachers, and other parties should get; and find suitable ways to communicate the impact of steering.

To address these limitations, we conducted a study on how adolescents trust an e-learning platform when they can steer recommended exercises and see their control's effects. Our research questions were as follows:

\begin{description}
\item[RQ1.] How does the ability to control recommended exercises affect students' trust in an e-learning platform?
\item[RQ2.] How is students' trust in an e-learning platform affected when they see a visual representation of their impact when controlling recommended exercises?
\end{description}

Our research contribution is threefold. First, we present a control mechanism and a visualisation of its impact, which have been found useful and usable by adolescents in a user-centred design process. Second, we discovered that a control mechanism does not necessarily change trust, neither when measured directly, nor when measured as a construct of competence, benevolence, integrity, intention to return, and perceived transparency. We also found, however, that a control mechanism can stimulate adolescents to reflect more upon their mastery level and the underlying recommendation system. Third, we show that visualising the control's impact can increase trust and perceived understanding of recommendations. Additionally, we share our dataset\footnote{\href{https://github.com/JeroenOoge/steering-recommendations-elearning}{https://github.com/JeroenOoge/steering-recommendations-elearning}} on how adolescents trust our platform and interact with our control mechanism, allowing further exploration and direct comparison in future research. In sum, our contributions highlight the potential of control mechanisms and related visualisations for adolescents in e-learning.


\section{Background and related work}
This section first discusses existing research on user control in recommender systems and then briefly highlights the overlap with explainable AI research, focusing on trust. Next, it zooms in on educational recommenders and relevant pedagogical background.

\subsection{Control over Recommendations}
In real-world settings, the accuracy of recommendation algorithms is subject to people's changing preferences: preference information known to the system can become outdated, leading to inaccurate recommendations~\cite{amatriain2009rate}. To ameliorate this problem, many control mechanisms have been developed to actively involve people in recommendation processes~\cite{jannach2017user}. For example, during preference elicitation, people can exercise control through preference forms~\cite{hijikata2012relation} or conversational dialogues~\cite{goker2000adaptive}. In addition, after being shown recommendation results, people can steer these results through critiquing~\cite{luo2020deep,petrescu2021multistep,chen2012critiquingbased}, dynamical filtering and re-sorting~\cite{bostandjiev2012tasteweights,odonovan2008peerchooser}, interactive (visual) explanations~\cite{he2016interactive,schaffer2015hypothetical,tsai2019explaining,tsai2021effects}, or changing the recommendation algorithms itself~\cite{ekstrand2015letting}.

Yet, how much control and which control mechanisms a recommender system should incorporate depends on the context, application, and end-user~\cite{cramer2008effects,jameson2002pros,jin2020effects}. Therefore, researchers have been studying the effects of providing control to end-users from different human-centred perspectives~\cite{konstan2012recommender,xiao2007ecommerce}, including perceived variety of recommendations, personal characteristics, trust, and understanding of the recommendation system. Specifically, Knijnenburg et al.~\cite{knijnenburg2012explaining} found that control can increase perceived variety of recommendations. Furthermore, preference for control methods in recommender systems depends on personal characteristics such as personality traits, need for cognition, and mood~\cite{knijnenburg2011each,jin2020effects,millecamp2018controlling}. Regarding trust in recommendations, control is highly valued for achieving personal goals but can also raise distrust about whether the control is just an illusion~\cite{harambam2019designing}. Finally, control mechanisms can increase overall system satisfaction and improve understanding of the recommendation process~\cite{knijnenburg2012inspectability}.


\subsection{Explainable AI and Trust}
The challenge to make recommendation algorithms more transparent fits in the wider field of \textit{explainable AI} (XAI). Essentially, XAI is an umbrella term for techniques that explain the outcomes of AI models, such that a specific audience can better understand and appropriately trust them~\cite{gunning2019darpa,barredoarrieta2020explainable,hind2019explaining,guidotti2019survey}. Research on these techniques brings together many concepts of interest, including fairness, privacy, bias, human reasoning, accountability, and ethics~\cite{abdul2018trends}.

One frequently studied concept in XAI is \textit{trust} in automated systems~\cite{lee2004trust}. Some work approaches trust from an algorithmic perspective, for example by considering it equivalent to reputation in recommender systems~\cite{odonovan2005trust}. However, XAI more often approaches trust from a human-centred perspective. Definitions for human-AI trust are heavily debated, but most agree that trust is an attitude in a situation of vulnerability and positive expectations~\cite{vereschak2021how}. Thus, from this angle, trust is a human belief that can be wrongly calibrated to the objective trustworthiness of an automated system~\cite{han2020trust}. Besides, trust building and calibration is influenced by how a system behaves: people's trust typically fluctuates until they feel sufficiently familiar with the system~\cite{yu2017user,nourani2020role,holliday2016user}.

Given the lack of well-accepted definitions, researchers measure human-AI trust in many ways. For example, some researchers consider trust as a \textit{one-dimensional}, i.e. monolithic, concept and typically measure it with a single Likert-type question. While some studies \aptLtoX[graphic=no,type=html]{[e.g.,40, 65]}{\citeg{holliday2016user,nourani2020role}} apply this strategy because it is quick, they are limited since a single question cannot measure a complex concept such as trust~\cite{hoff2015trust}. Alternatively, other researchers consider trust as a \textit{multidimensional} ensemble of several constructs which they typically measure with multiple Likert-type questions. For example, \mbox{McKnight} et al.~\cite{mcknight2002developing} introduced \textit{trusting beliefs} as a composition of competence, benevolence, and integrity; and Ooge et al.~\cite{ooge2022explaining} measured trust as the average of trusting beliefs, intention to return, and perceived transparency.


\subsection{Educational Recommender Systems}
Recommendation techniques are increasingly being integrated in digital learning environments~\cite{zhai2021review,khanal2020systematic}. However, educational recommender systems differ from their general-purpose counterparts: they intend to facilitate achieving learning goals, are subject to a pedagogical context, and consider end-users' educational role or mastery level instead of personal characteristics~\cite{garcia-martinez2013educational,manouselis2013recommender}. In general, educational recommender systems can support learning in several ways~\cite{drachsler2015panorama}. For example, they can recommend courses~\cite{farzan2011encouraging,aher2013combination}, suggest additional learning resources~\cite{tang2005smart}, and support teachers to improve their courses or monitor their teaching resources~\cite{gallego2013enhanced,garcia2009architecture}.

In the spirit of XAI for education~\cite{khosravi2022explainable}, educational recommender systems are often requested to allow steering and to justify their recommendations. Steering could occur, for example, in the form of explicitly asking learners for feedback on exercises' difficulty after completing them~\cite{michlik2010exercises}. Furthermore, recommendations tailored to learners' mastery level can be justified by showing how the system estimates that mastery level~\cite{kay2019data}. In the context of open learner models~\cite{bull2010open,conati2018ai,hooshyar2020open}, Mabbott and Bull~\cite{mabbott2006student} found that learners felt less comfortable having full control over a learner model, compared to only making suggestions; and Abdi et al.~\cite{abdi2020complementing} found that an open learner model increases understanding of recommendations.


\subsection{Estimating Mastery and Exercise Difficulty}
From a pedagogical perspective, students' mastery level can be assessed based on several frameworks. One famous framework is Bloom's revised taxonomy~\cite{krathwohl2002revision}, which consists of two dimensions: a knowledge dimension with four levels (factual, conceptual, procedural, and metacognitive knowledge) and a cognitive process dimension with six levels (remember, understand, apply, analyse, evaluate, and create). Another framework is the Dreyfus model~\cite{dreyfus2004fivestage}, which proposes five skill acquisition levels: novice, advanced beginner, competent, proficient, and expert. 

From a computer science perspective, different techniques can simultaneously estimate learners' mastery level and exercises' difficulty based on how learners perform while solving exercises~\cite{wauters2012item,torkamaan2022recommendations}. Specialised models such as item response theory~\cite{kadengye2015modeling} or knowledge tracing~\cite{guo2021enhancing}, however, need to be calibrated on large item sets with known difficulties~\cite{wauters2012item,pelanek2016applications}. A classic alternative that circumvents this disadvantage is the \textit{Elo rating system}~\cite{pelanek2016applications}, which was originally introduced by Arpad Elo~\cite{elo1978rating} for rating chess players. Translated to education, the Elo rating system assigns dynamic ratings to both learners and exercises: the higher a learner's rating, the higher their mastery level; and the higher an exercise's rating, the more difficult it is. Furthermore, Elo ratings are of interval scale and their range can be chosen arbitrarily. Each time a learner $l$ answers an exercise $e$, the Elo ratings of $l$ and $e$ are updated as follows:
\begin{align}
 \elo(l)&=\elo(l)+k\cdot (X_{le}-P(X_{le}=1))\nonumber\\
 \text{and}\quad
 \elo(e)&=\elo(e)-k\cdot (X_{le}-P(X_{le}=1)),
    \label{eq:elochanges}
\end{align}
where $k$ is a fixed learning-rate parameter that determines how strongly the attempt influences the Elo rating, $X_{le}\in\{0,1\}$ reflects whether $l$ answered $e$ correctly, and
\begin{equation}
P(X_{le}=1)=1/\big(1+\exp({\elo(e)-\elo(l)})\big)
\label{eq:eloprob}    
\end{equation}
is the modelled probability for a correct answer. In words, whenever someone correctly solves an exercise, their Elo rating increases and the exercise's Elo rating decreases, proportional to how unexpected that correct answer was; vice versa for incorrect answers. Besides its intuitive functioning, the Elo rating system has the asset that it can be extended to multivariate settings~\cite{abdi2019multivariate}, adapted to consider how quickly students solve questions~\cite{klinkenberg2011computer}, and combined with other techniques such as collaborative filtering~\cite{dahl2018combining,ooge2022explaining}.



\section{Materials and Methods}
This section presents our e-learning platform and design decisions inspired by a pilot study with teachers and an iterative design process with students. Next, it describes our main study design, which was approved by the ethical committee of \anon{KU Leuven} (reference number: \anon{G-2022-4917}).


\subsection{E-Learning Platform with Personalised Exercises and a Control Mechanism}
We built upon \textit{Wiski}, an existing e-learning platform for middle and high school students~\cite{ooge2019personaliseren}. Essentially, Wiski's core functionality is solving multiple-choice questions about maths topics in the Belgian school curriculum. Through an iterative design process with students and teachers, we extended this core with three functionalities: (a)~composing exercise series recommended for students' mastery level, (b)~giving students partial control over their estimated mastery level, and (c)~visualising the impact of that control. Think-aloud studies in which adolescents executed predefined tasks on a low-fidelity version of our e-learning platform ensured that these new functionalities were deemed useful and usable. 

\begin{figure}
  \centering
  \includegraphics[width=\linewidth]{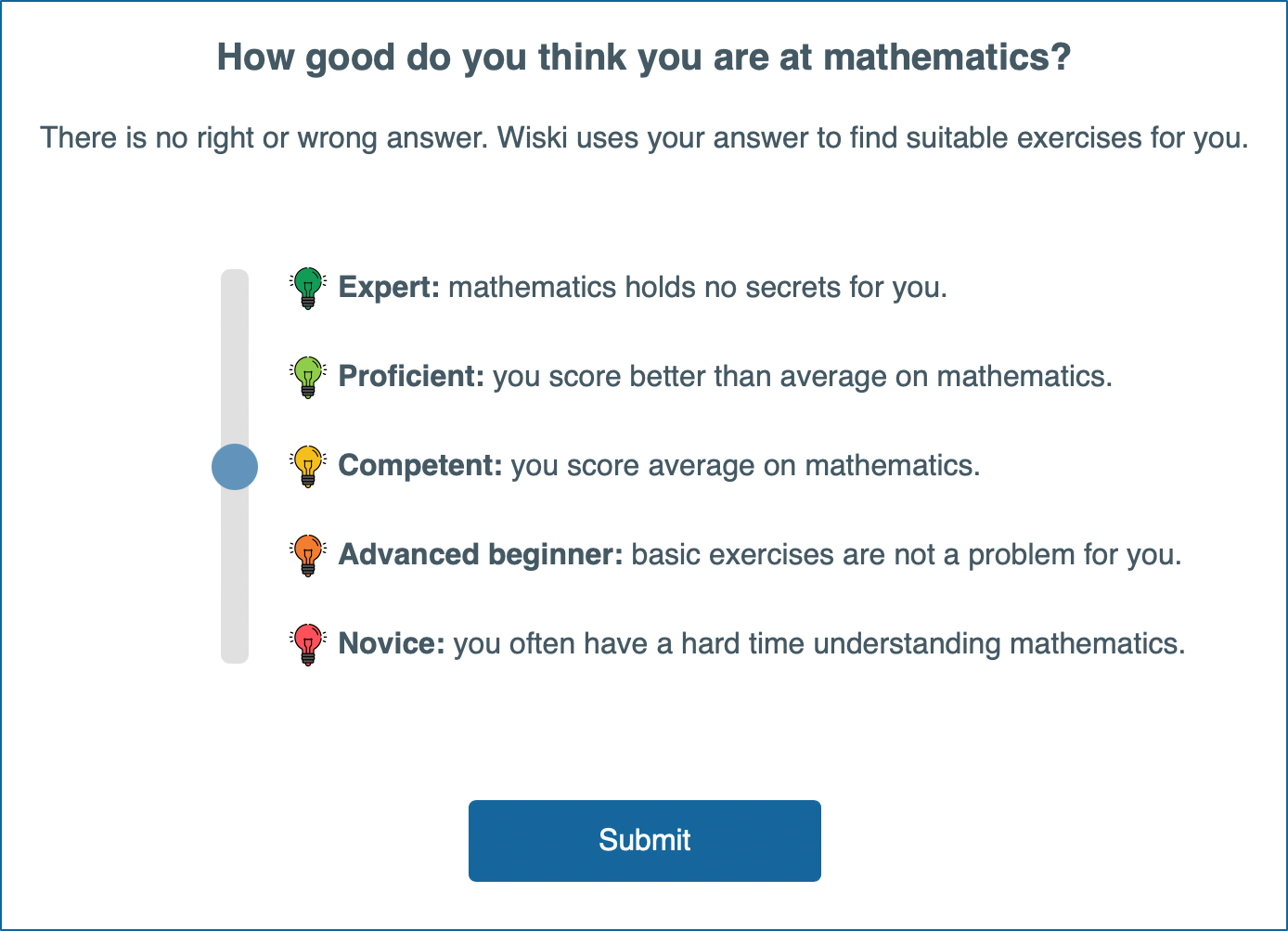}
  \captionof{figure}{Students initialised their maths mastery level with a continuous slider that indicated five thresholds: novice, advanced beginner, competent, proficient, and expert.}
  \label{fig:initialmastery}
\end{figure}

\paragraph{Personalised exercise series}
Brief semi-structured interviews~\cite{leech2002asking} with 4 high school teachers learned us that teachers appreciated the idea of an e-learning platform that recommends exercises tailored to students' mastery level. In addition, to give students sufficient time to adapt to new difficulty levels, teachers advised recommending exercise \textit{series} instead of individual exercises. We therefore decided to let our platform estimate students' mastery level and exercises' difficulty with an Elo rating system and then use those estimates to recommend exercise series. Specifically, whenever a student $l$ would select a topic to practise, they would start a series consisting of two exercises, $e_1$ and $e_2$, chosen such that $P(X_{le_1}=1)$ and $P(X_{le_2}=1)$ were closest to 0.7; a value yielding reasonably challenging exercises~\cite{klinkenberg2011computer}. Probabilities were estimated with a variant of (\ref{eq:eloprob}), which originates from a chess context:
\[
P(X_{le}=1)=1/\big(1+10^{(\elo(e)-\elo(l))/400}\big).
\]
To set up our Elo rating system, students could initialise their Elo rating with the slider in \Cref{fig:initialmastery}, which indicated five thresholds inspired by the Dreyfus model~\cite{dreyfus2004fivestage}. In the background, the slider's range corresponded to the interval $[1000, 2000]$, which roughly corresponds to typical Elo scores for novice~(1000) and expert~(2000) chess players. Furthermore, exercises' initial Elo ratings were set by teachers who participated in our main study. Concretely, teachers used the thresholds in \Cref{fig:initialmastery} to estimate the difficulty of all exercises belonging to the subjects they wanted to cover in class. In case multiple teachers were interested in the same subjects, we only asked one of them to set the initial ratings, distributing the workload evenly. Finally, we set the hyperparameter $k$ in (\ref{eq:elochanges}) to 160 to allow for relatively large Elo changes.

\aptLtoX[graphic=no,type=html]{\begin{figure*}
\begin{minipage}[b]{.48\linewidth}
  \centering
  \includegraphics[width=\linewidth]{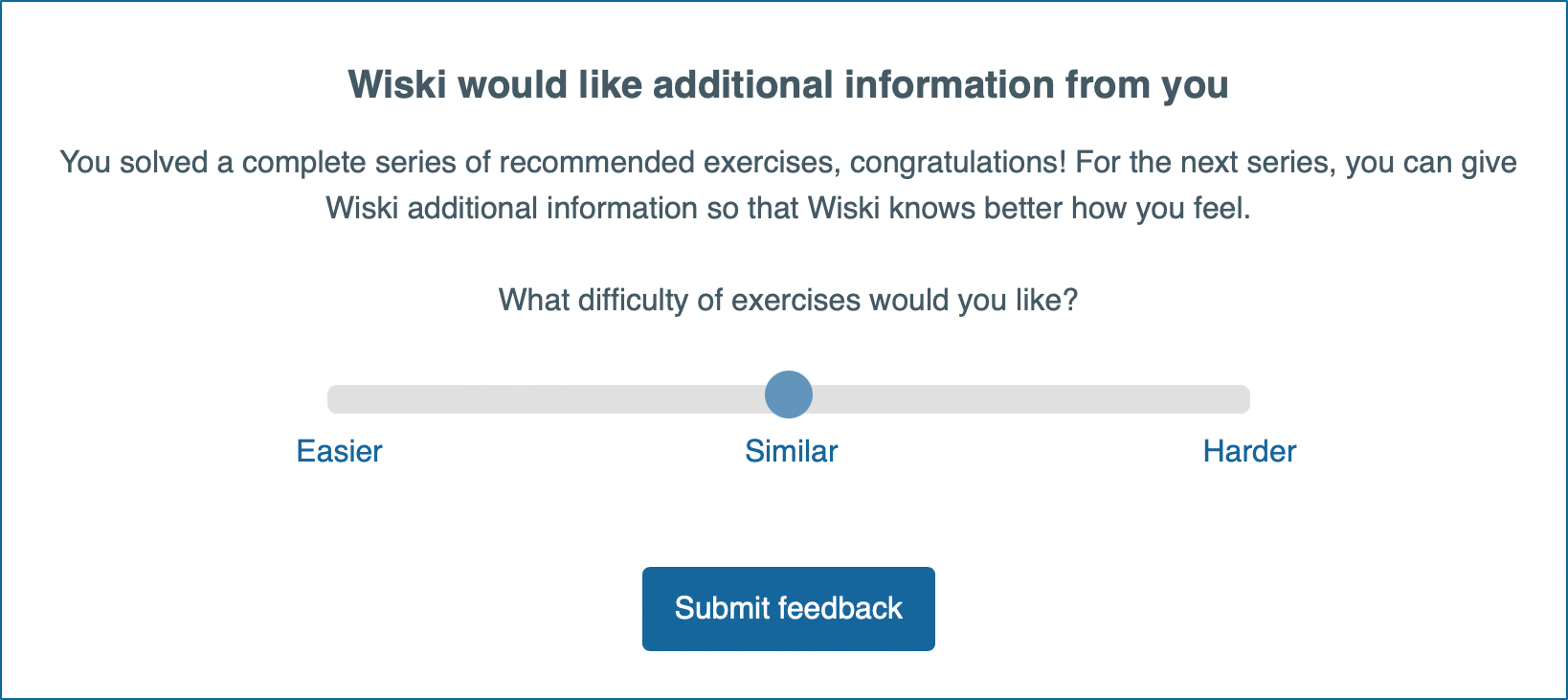}
  \captionof{figure}{After each exercise series, students could steer subsequent recommendations with a 20-step slider: lowering their mastery level yielded easier series, and vice versa.}
  \label{fig:control}
\end{minipage}
\end{figure*}
\begin{figure*}
  \begin{minipage}[b]{.48\linewidth}
  \includegraphics[width=\linewidth]{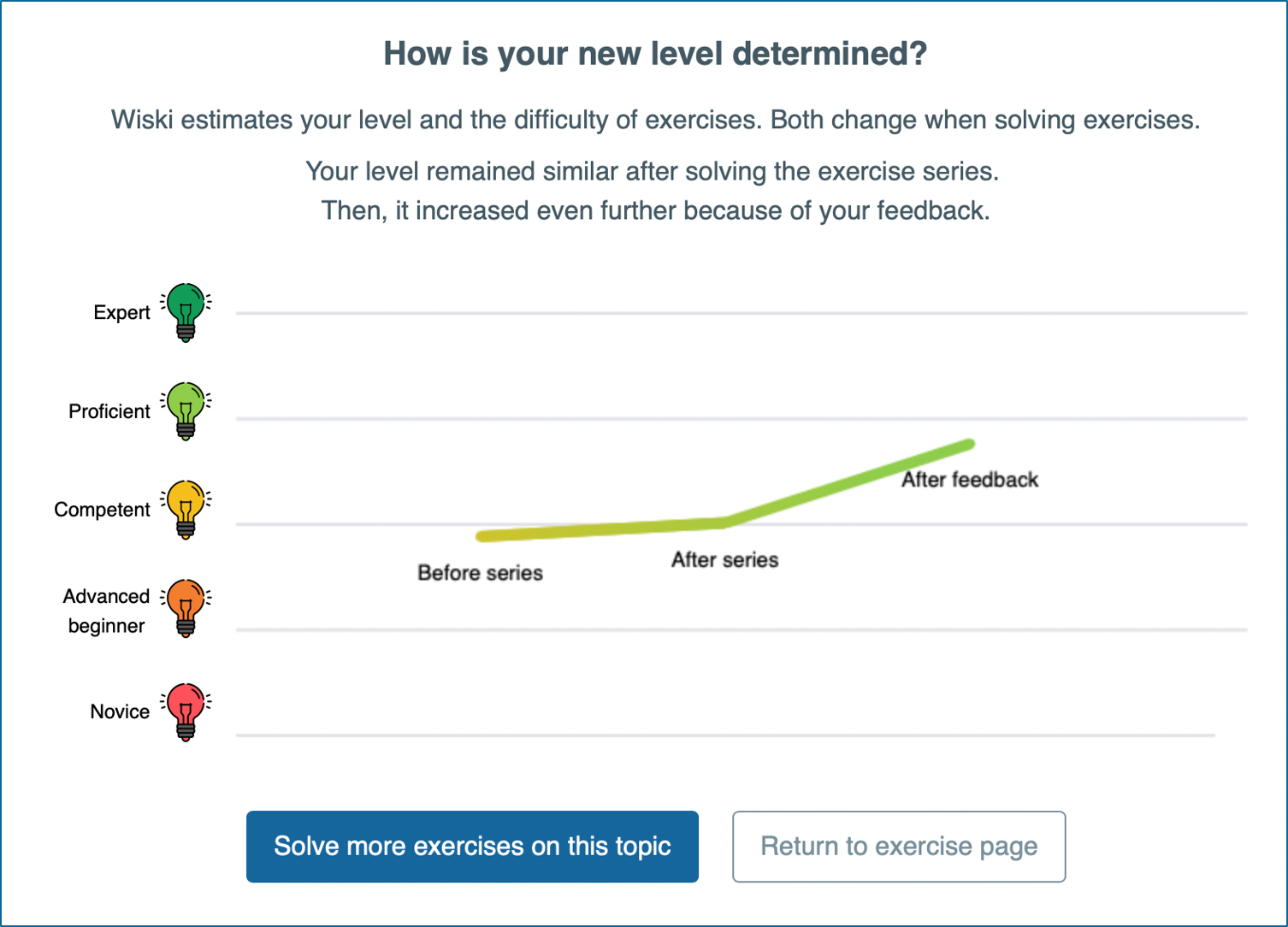}
  \captionof{figure}{Visualisation of students' steering impact after an exercise series. The top describes the evolution of students' mastery level; the bottom visualises it.}
  \label{fig:impact}
  \end{minipage}
\end{figure*}
}{
\begin{figure*}
\begin{minipage}[b]{.48\linewidth}
  \centering
  \includegraphics[width=\linewidth]{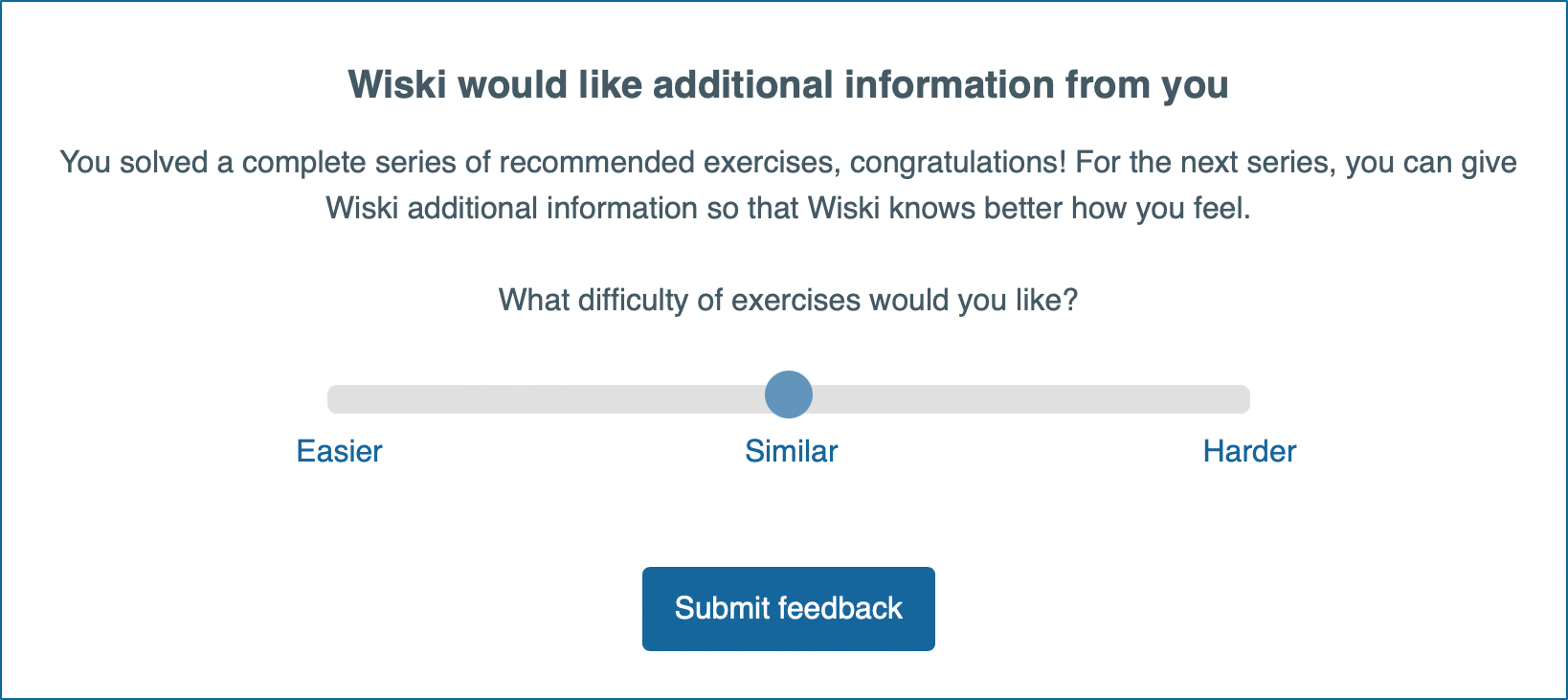}
  \captionof{figure}{After each exercise series, students could steer subsequent recommendations with a 20-step slider: lowering their mastery level yielded easier series, and vice versa.}
  \label{fig:control}
\end{minipage}\hfill%
  \begin{minipage}[b]{.48\linewidth}
  \includegraphics[width=\linewidth]{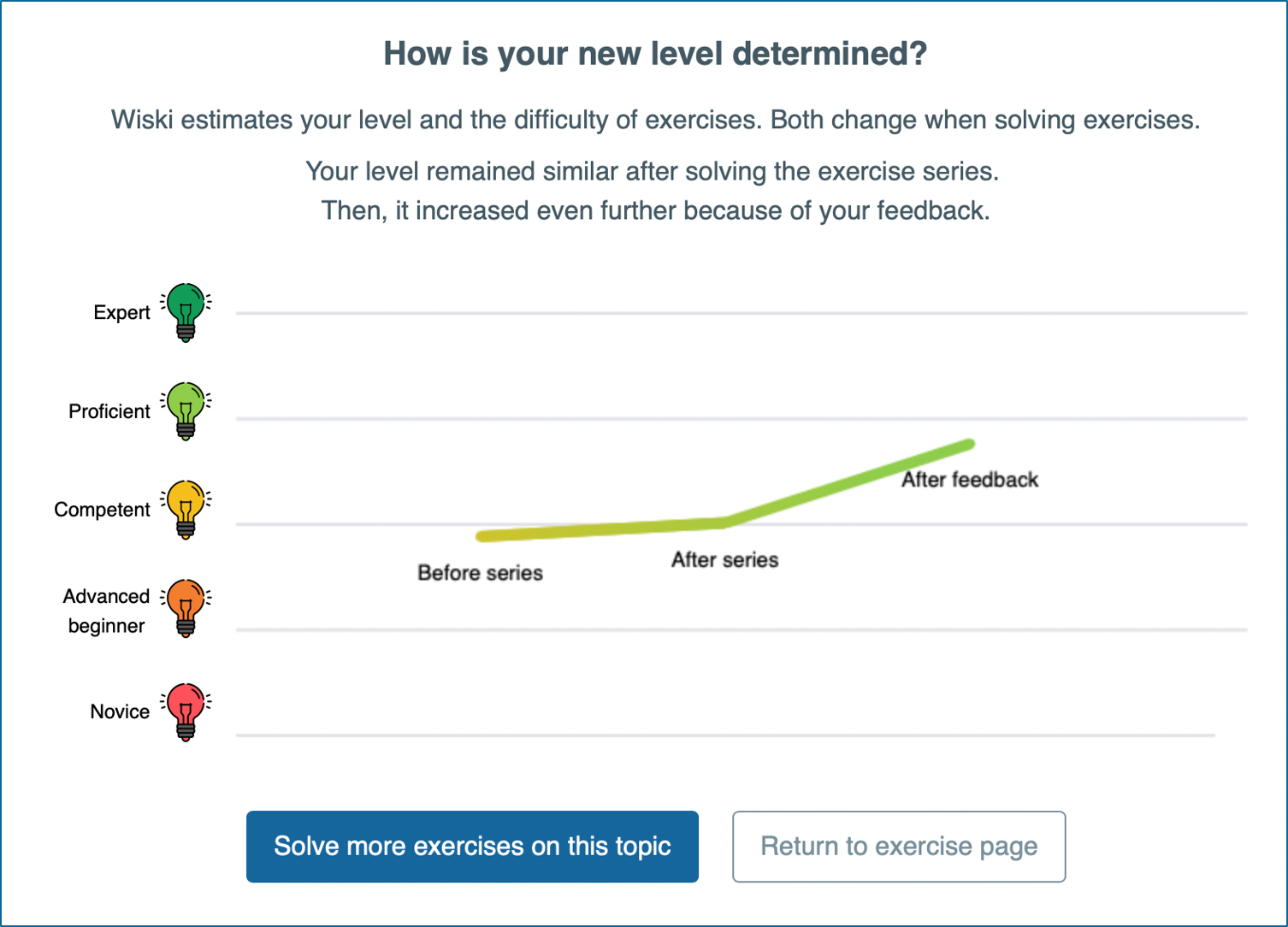}
  \captionof{figure}{Visualisation of students' steering impact after an exercise series. The top describes the evolution of students' mastery level; the bottom visualises it.}
  \label{fig:impact}
  \end{minipage}
\end{figure*}}

\paragraph{Control mechanism and impact visualisation}
Through two rounds of think-aloud studies with 11~adolescents (2~middle school, 9~high school), we iteratively designed a control mechanism and a visualisation of the exercised control's impact. First, the control mechanism in \Cref{fig:control} allowed students to modify their mastery level and thus steer the difficulty of subsequent recommendations. Specifically, after finishing an exercise series, students could indicate whether they wanted easier or harder exercises. In the background, this would lower or raise their Elo rating up to 10\%, respectively. The think-alouds learned us that the slider provided intuitive and sufficient control. In addition, adolescents preferred to reflect in terms of their mastery level and were sometimes confused by a preliminary design that also allowed them to steer exercises' Elo ratings with a similar slider. Second, the visualisation of the control's impact in \Cref{fig:impact} contained three parts: a fixed explanation; a description of how mastery level changed due to solving an exercise series and subsequent steering; and a line chart of the latter information. The think-alouds learned us that adolescents preferred the line chart over an animated bar chart. More details on our iterative designs can be found in \anon{Dereu's} master's thesis \anon{\cite{dereu2022juiste}}.

\subsection{Study Design}
\begin{figure*}
  \centering
  \includegraphics[width=0.95\linewidth]{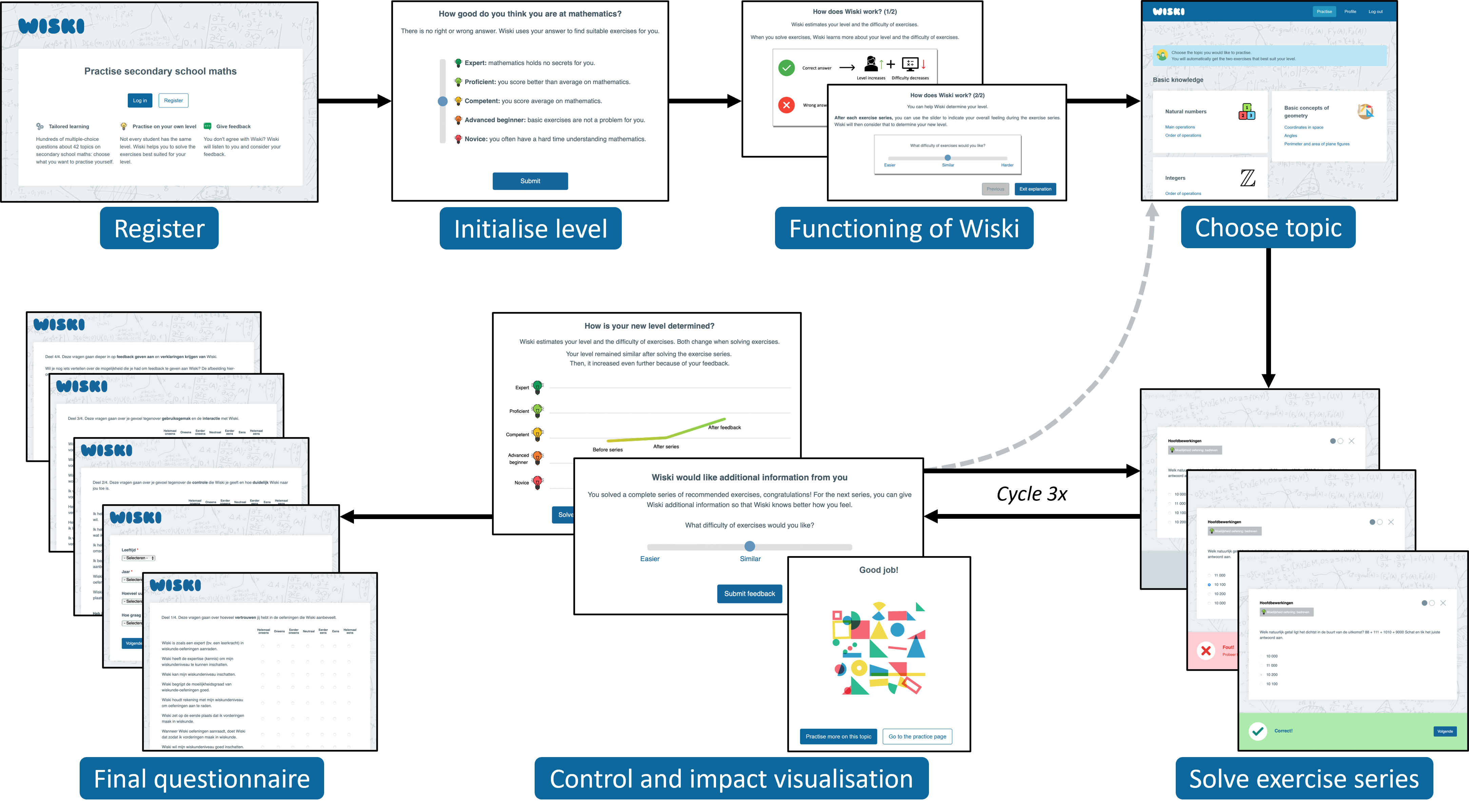}
  \caption{Flow of our study: registering, picking an initial mastery level, reading a global explanation on the functioning of Wiski, solving three series (i.e., six exercises) potentially followed by steering one's mastery level and seeing its impact, and finally filling out a questionnaire.}
  \label{fig:flow}
\end{figure*}

\begin{figure*}
  \centering
  \vspace{\baselineskip}
  \includegraphics[width=.85\linewidth]{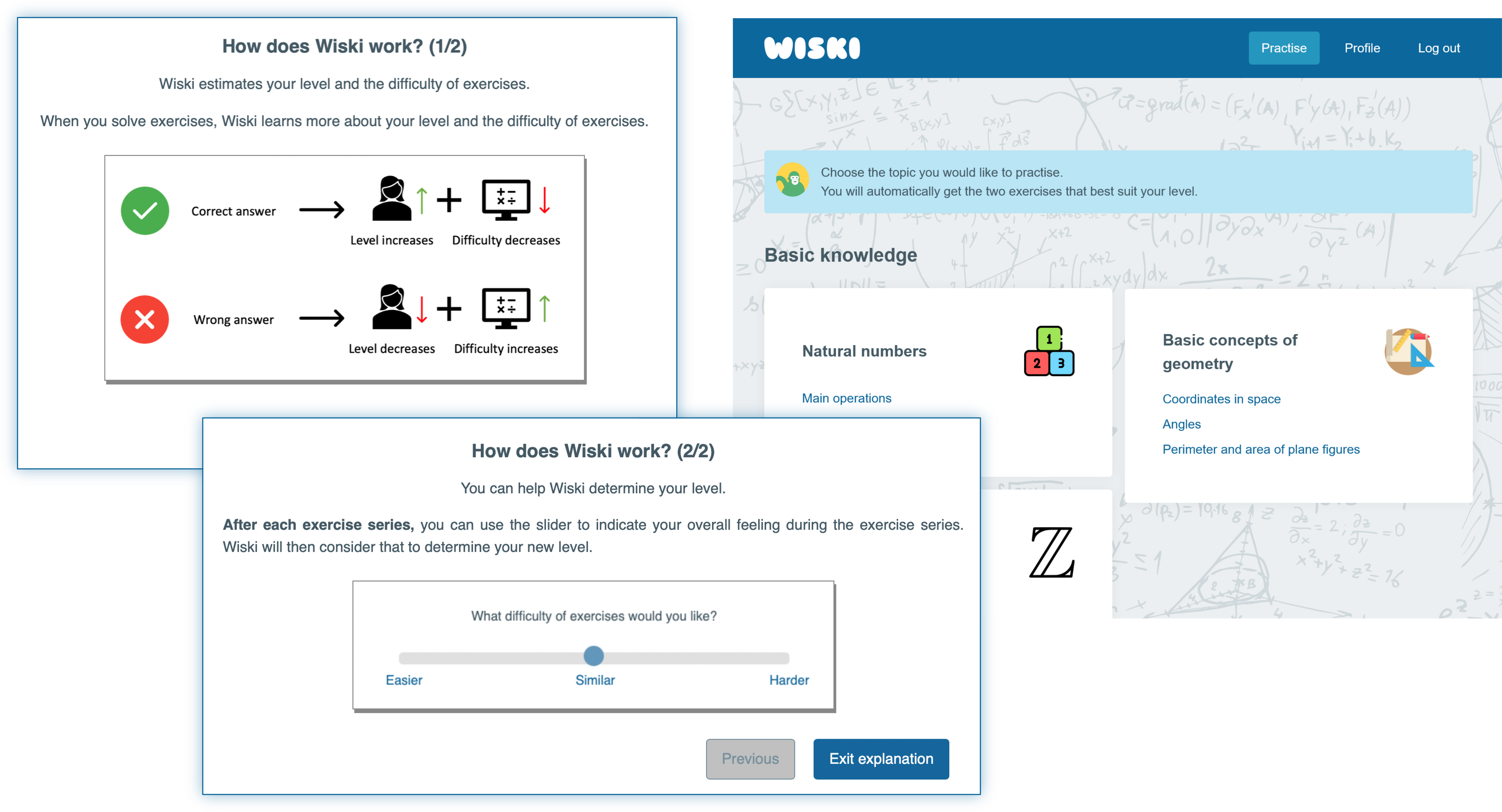}
  \caption{Wiski explained in two ways how it personalises exercise series. (l) After registration, all participants saw a global explanation; participants in \textsc{control} and \textsc{control+impact} saw an additional screen. (r) The practice page for picking maths topics explained recommendations: \quote{You will automatically get the two exercises that best suit your level.}}
  \label{fig:explanations}
\end{figure*}

To answer our research questions, we conducted a randomised controlled experiment with three groups: in \textsc{none}, participants did not have any control over recommended exercises; in \textsc{control}, participants could steer their mastery level with the slider in \Cref{fig:control}; and in \textsc{control+impact}, participants additionally saw the visualisation of their control's impact in \Cref{fig:impact}. The flow of our study is depicted in \Cref{fig:flow}. First, participants registered on our Wiski platform and were randomly assigned to one of the three research groups. Then, they initialised their maths mastery level with the slider in \Cref{fig:initialmastery} and saw one or two of the screens in \Cref{fig:explanations} which globally explained Wiski's recommendation algorithm. Next, participants chose a maths topic on the practice page and solved three series, each consisting of two exercises. We chose this relatively low number of series to ensure participants could finish the study in under 50~minutes. After each series, participants could adjust their mastery level and see its impact, depending on their research group. Finally, participants filled out a questionnaire and could continue to freely use the platform. Thus, participants' experience with Wiski only differed in whether or not they could control their mastery level and see a visualisation of their control's impact. In the background, we logged all Elo rating changes.

Our final questionnaire contained the 31 Likert-type questions in \Cref{tab:questionnaire}, scored on a 7-point range. The first part captured trust. Similar to Ooge et al.~\cite{ooge2022explaining}, we measured trust both with a single question and as the average of trusting beliefs, intention to return, and transparency. Slightly different is that, for more reliable scores, we measured transparency with three questions from the ResQue questionnaire~\cite{pu2010usercentric} instead of one. The second part of our questionnaire, also based on ResQue~\cite{pu2010usercentric,pu2011usercentric}, captured three control aspects: overall control, preference elicitation, and preference revision.

Our questionnaire also contained open text fields that encouraged participants to elaborate on their Likert-type responses. Furthermore, we explicitly asked participants whether they trusted our platform for recommending maths exercises, whether they (would have) liked controlling the desired difficulty level of exercises, and whether they (would have) liked seeing the impact of that control. Only the open question on trust was mandatory and the latter two questions included screenshots similar to \Cref{fig:control,fig:impact}.

\subsection{Participant Recruitment}
We contacted 30 secondary school teachers in \anon{Belgium (Flanders)} via email and LinkedIn, inviting them and their students to participate in our research during school hours. We asked teachers to not coerce students into participating and to prepare exercises on paper should some students refuse to participate. Four teachers accepted our invitation: they passed through a brochure to students and their respective parents, which communicated our study goals, data management, and Covid-19 precautions. Interested students gave informed consent and students under 16 required parental consent. Ultimately, all 76 invited students (ages 12–17) participated in the study. We excluded 5 participants from the analysis due to incomplete questionnaires, ending up with 22 participants in \textsc{none}, 25 in \textsc{control}, and 24 in \textsc{control+impact}.

\begin{figure}
    \centering
    \includegraphics[width=.9\linewidth]{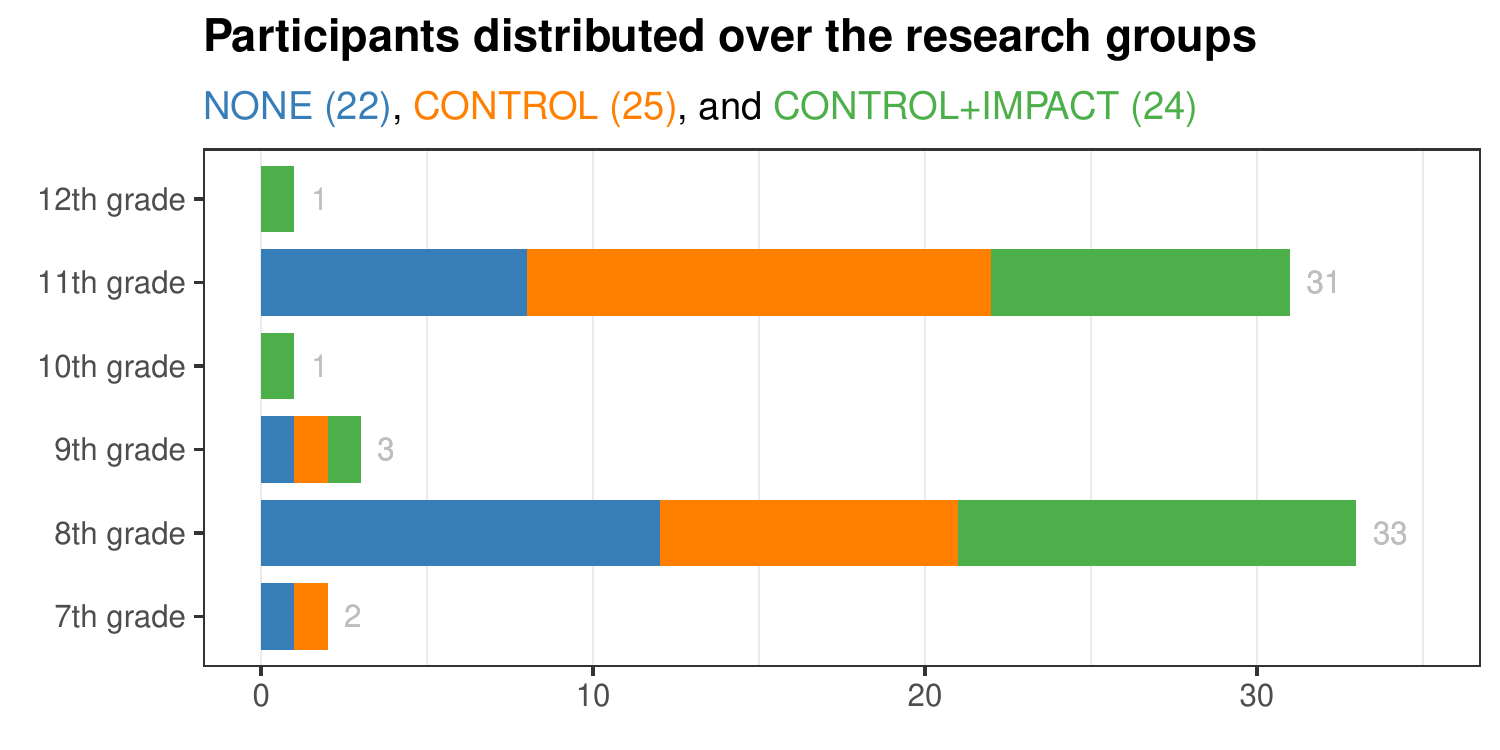}
    \captionof{figure}{Participants distributed over the research groups per grade: most were in 8th and 11th grade.}
    \label{fig:participants}
\end{figure}

\begin{table}[t]
\caption{Comparing the research groups with t-tests (Mann-Whitney U test for one-dimensional trust). Cells contain the effect sizes (second group mean minus first group mean).}
	\label{table:tests}
	\centering\small
	\def\colwidth{2cm}
\resizebox{1\linewidth}{!}{%
	\begin{tabular}{@{}lp{50pt}p{55pt}p{60pt}@{}}
		\toprule
		                       & \textsc{none} \textbf{vs.} \newline \textsc{control}     & \textsc{none} \textbf{vs.}\newline \textsc{control+impact} & \textsc{control} \textbf{vs.}\newline \textsc{control+impact} \\
		\midrule\addlinespace[1ex]
		Benevolence            & \insignif{\phantom{-}0.16 \scriptsize{($p=0.263$)}}                          & \phantom{-}0.61 \scriptsize{($p=0.011$)}   & \phantom{-}0.45 \scriptsize{($p=0.035$)}            \\
		Trusting beliefs       & \insignif{-0.01 \scriptsize{($p=0.529$)}}                                    & \phantom{-}0.38 \scriptsize{($p=0.042$)}   & \phantom{-}0.40 \scriptsize{($p=0.030$)}            \\
		Transparency           & \insignif{\phantom{-}0.29 \scriptsize{($p=0.068$)}}                          & \phantom{-}1.04 \scriptsize{($p=0.000$)}** & \phantom{-}0.74 \scriptsize{($p=0.002$)}*           \\
		One-dimens. trust  & \insignif{\phantom{-}0.00 \scriptsize{($p=0.504$)}}                          & \phantom{-}0.78 \scriptsize{($p=0.017$)}   & \phantom{-}0.78 \scriptsize{($p=0.020$)}            \\
		Multidimens. trust & \insignif{\phantom{-}0.15 \scriptsize{($p=0.207$)}}                          & \phantom{-}0.55 \scriptsize{($p=0.009$)}*  & \phantom{-}0.40 \scriptsize{($p=0.039$)}            \\
		Preference revision    & \insignif{\phantom{-}0.33 \scriptsize{($p=0.080$)}}                          & \phantom{-}0.43 \scriptsize{($p=0.030$)}   & \insignif{\phantom{-}0.10 \scriptsize{($p=0.325$)}} \\
		\bottomrule
		\multicolumn{4}{@{}l}{\footnotesize{*$p<0.01$, **$p<0.001$, non-significant results ($p\geq 0.5$) are greyed out}}
	\end{tabular}}
\end{table}

\subsection{Data Analysis}
We analysed the collected quantitative data in R~4.2.1. To compare the three research groups in terms of trust and control perceptions, measured as an \textit{average} of several Likert-type questions, we first conducted one-way ANOVA tests after checking the requirements: independence was guaranteed by the randomised set-up, assuming a normal distribution was plausible given the central limit theorem, and equal variances were verified with F-tests. Then, constructs that differed in at least two groups ($p<0.10$) were compared in more detail with unpaired t-tests, which assume the same as ANOVA. To compare trust measured with a \textit{single} Likert-type question, we applied Mann-Whitney U tests to avoid normality assumptions. All t-tests and Mann-Whitney U tests used $p<0.05$ as threshold for significance and were one-sided with alternative hypothesis that groups with more functionalities score higher.

To get further insights into differences between research groups, we thematically analysed \cite{braun2012thematic} the qualitative feedback stemming from the open questions in our questionnaire. For presentation here, we translated the original Dutch responses to English, only correcting grammar and spelling.

\section{Results}
\begin{figure*}
  \centering
  \includegraphics[width=\linewidth]{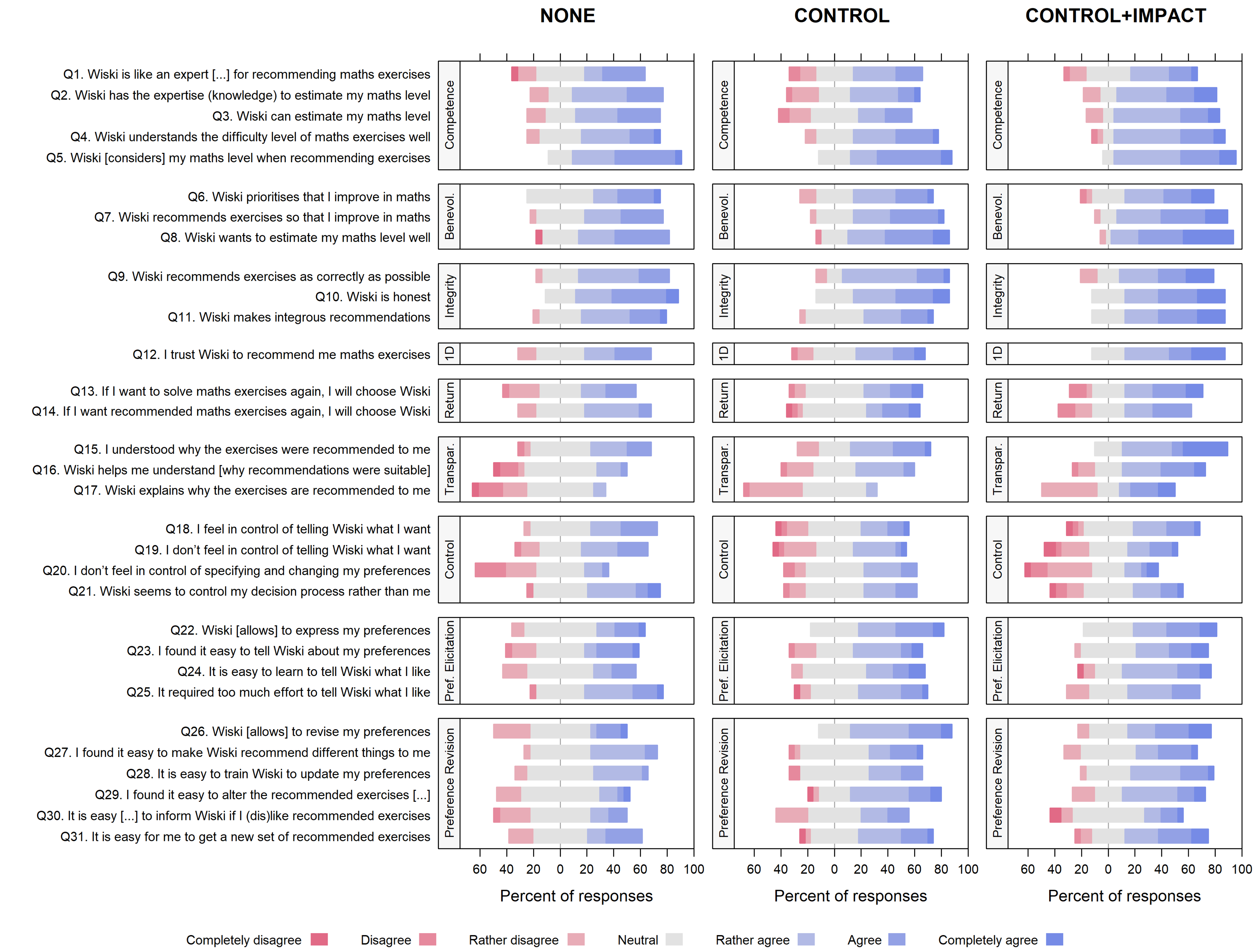}
  \caption{Diverging bar charts \cite{heiberger2014design} of responses to the questionnaire in \Cref{tab:questionnaire} after reverse-scoring, comparing the three research groups. Questions have been abbreviated for brevity and have been grouped per construct for clarity.}
  \label{fig:divergingBarchart}
\end{figure*}

\begin{figure*}
    \centering
    \begin{minipage}{.85\linewidth}
        \includegraphics[width=.24\textwidth]{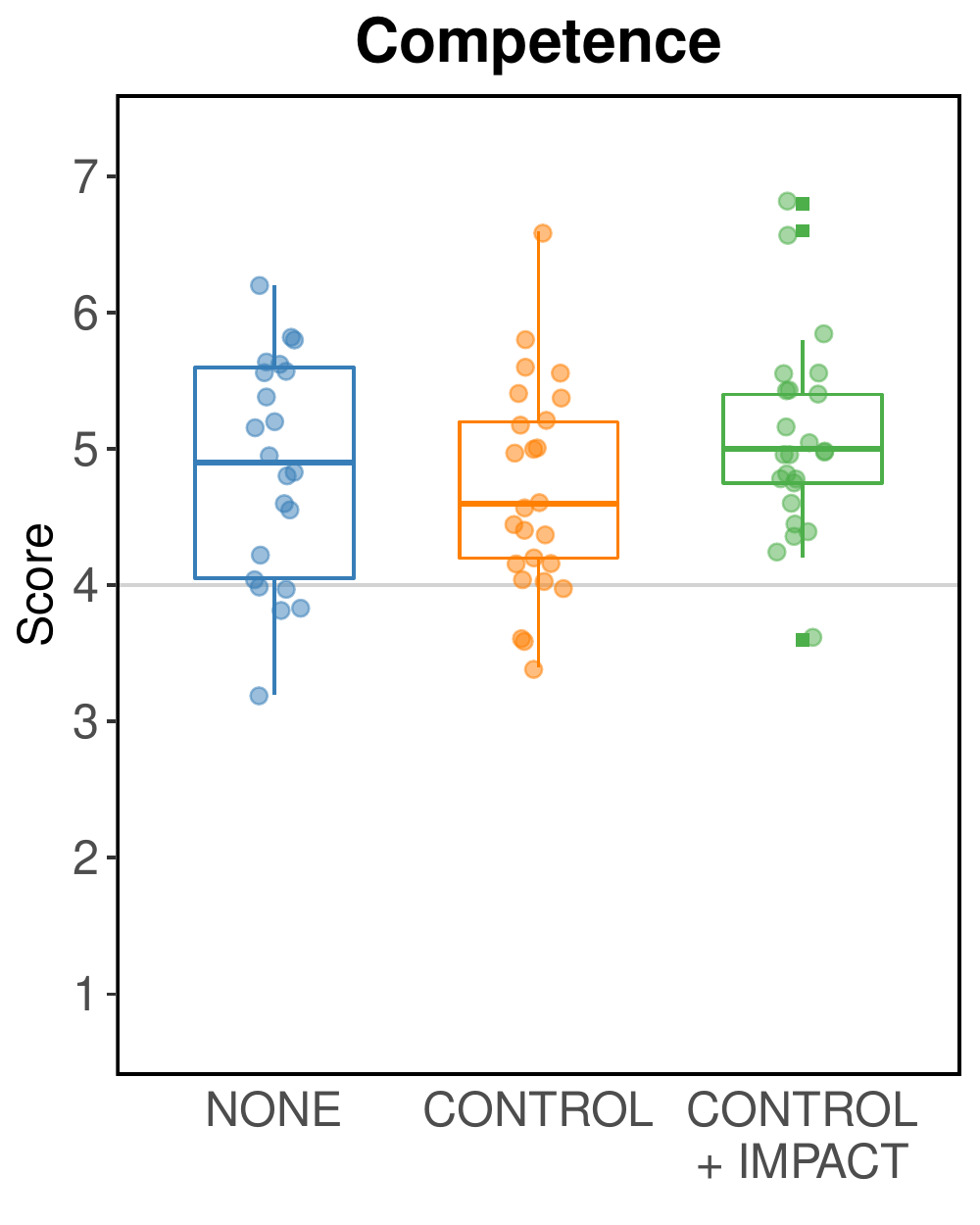}\hfill
        \includegraphics[width=.24\textwidth]{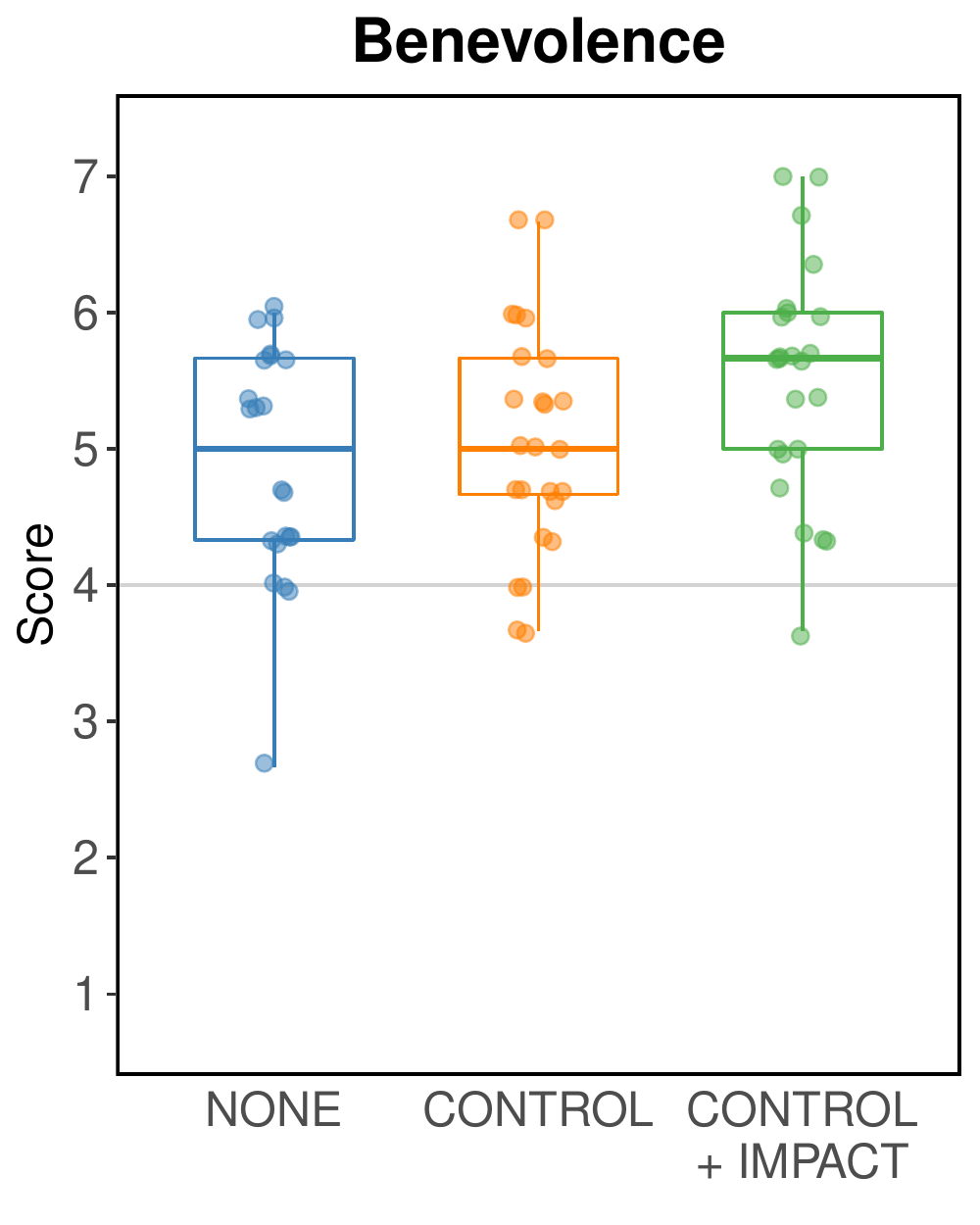}\hfill
        \includegraphics[width=.24\textwidth]{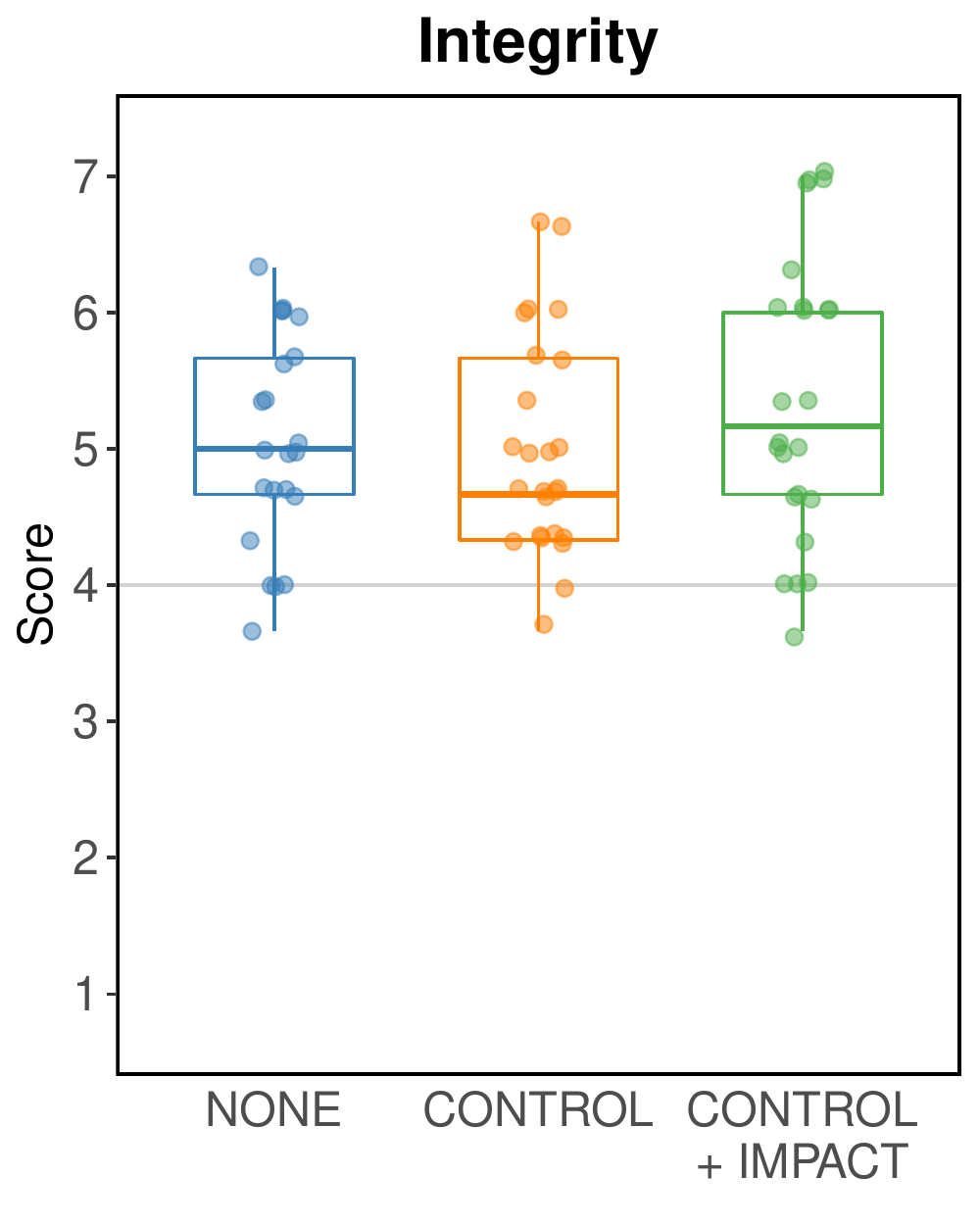}\hfill
        \includegraphics[width=.24\textwidth]{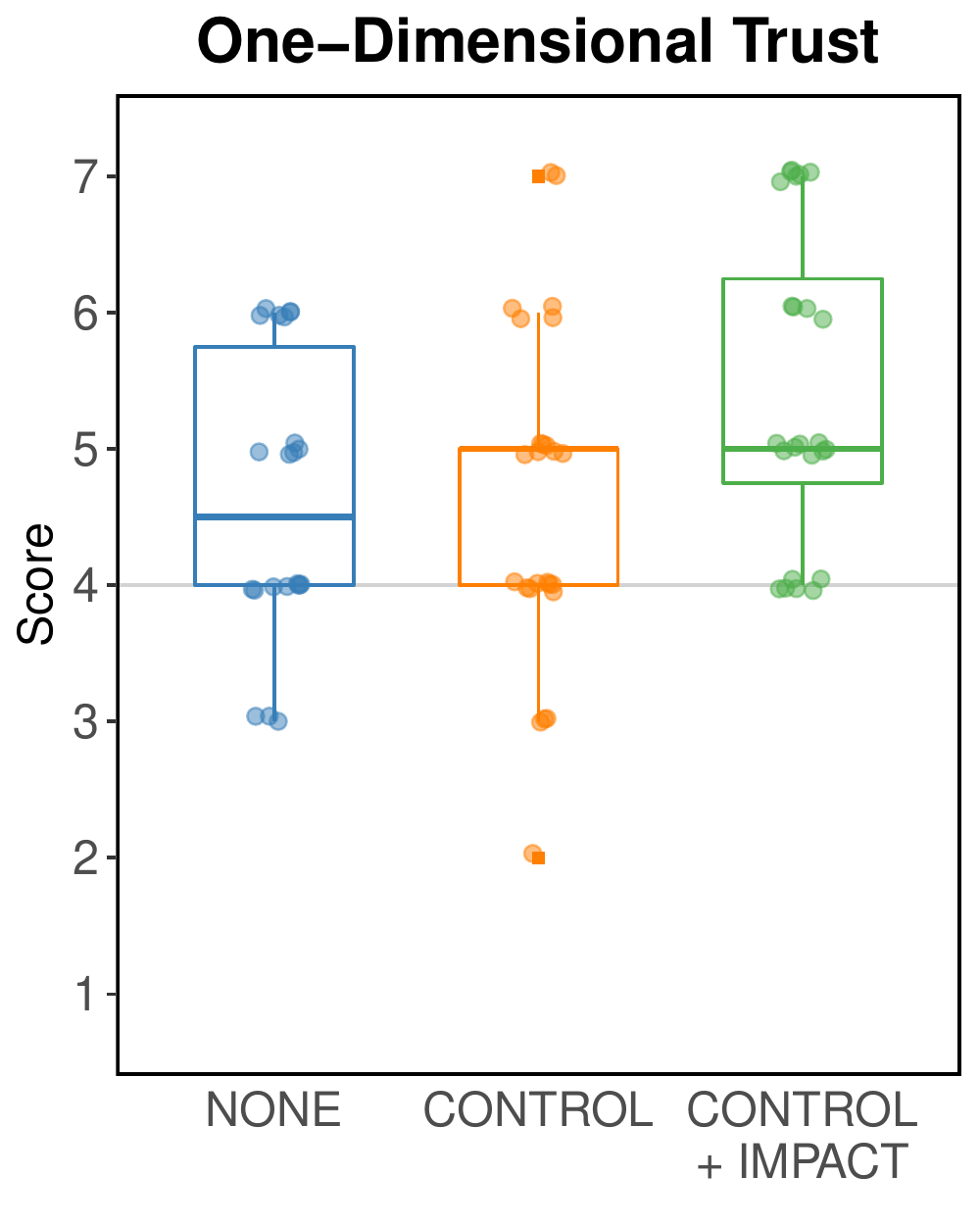}
    \end{minipage}
    \begin{minipage}{.85\linewidth}
        \includegraphics[width=.24\textwidth]{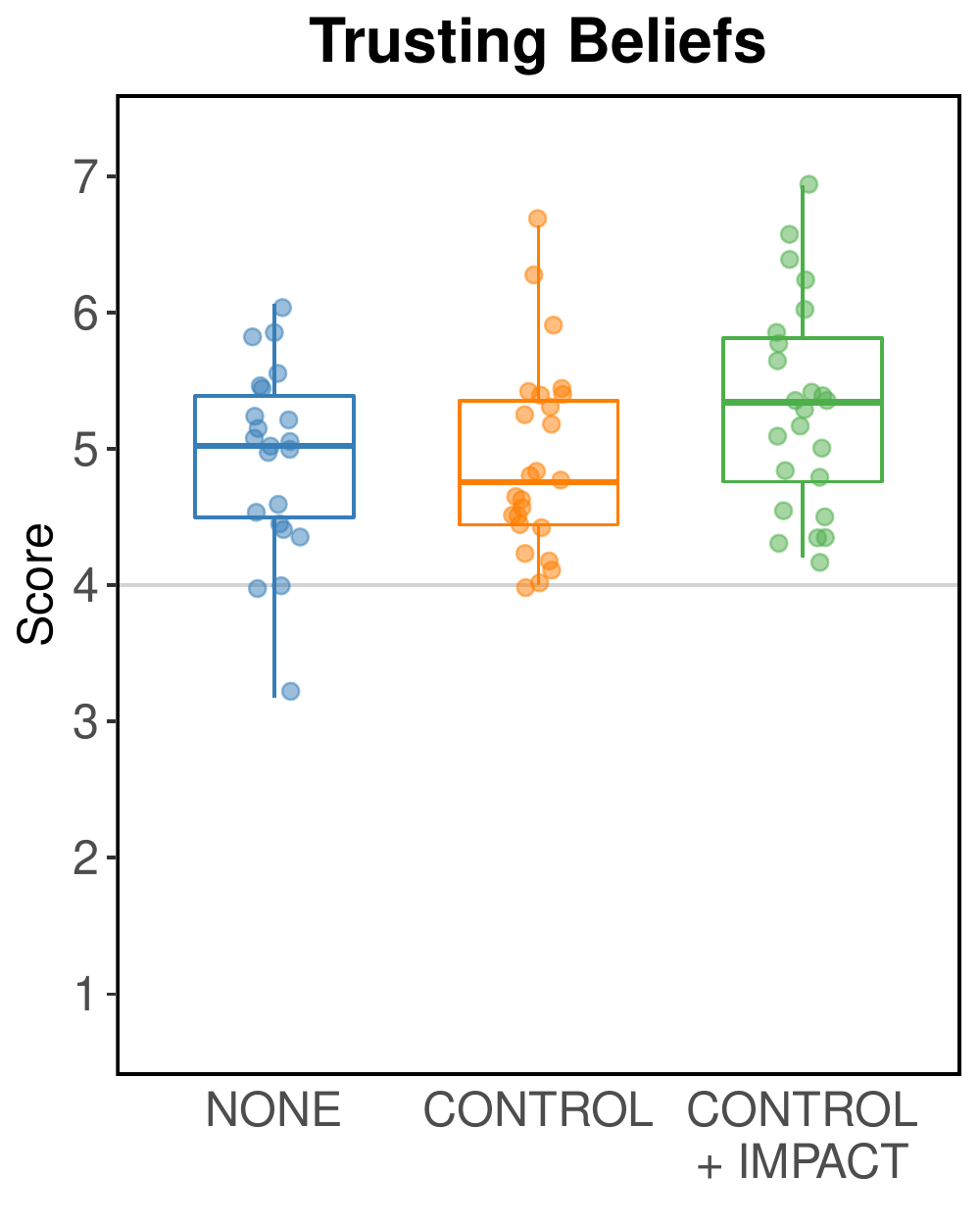}\hfill
        \includegraphics[width=.24\textwidth]{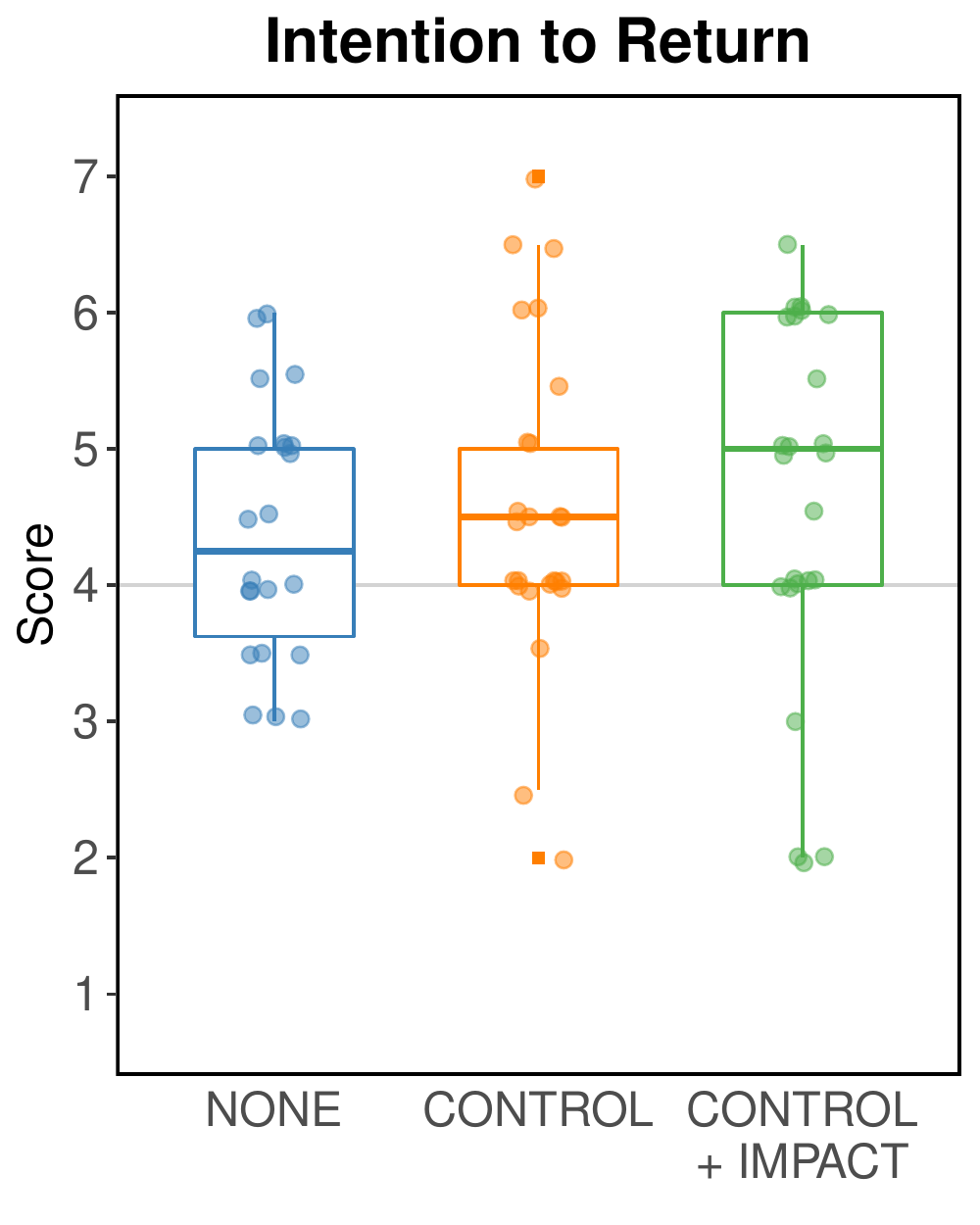}\hfill
        \includegraphics[width=.24\textwidth]{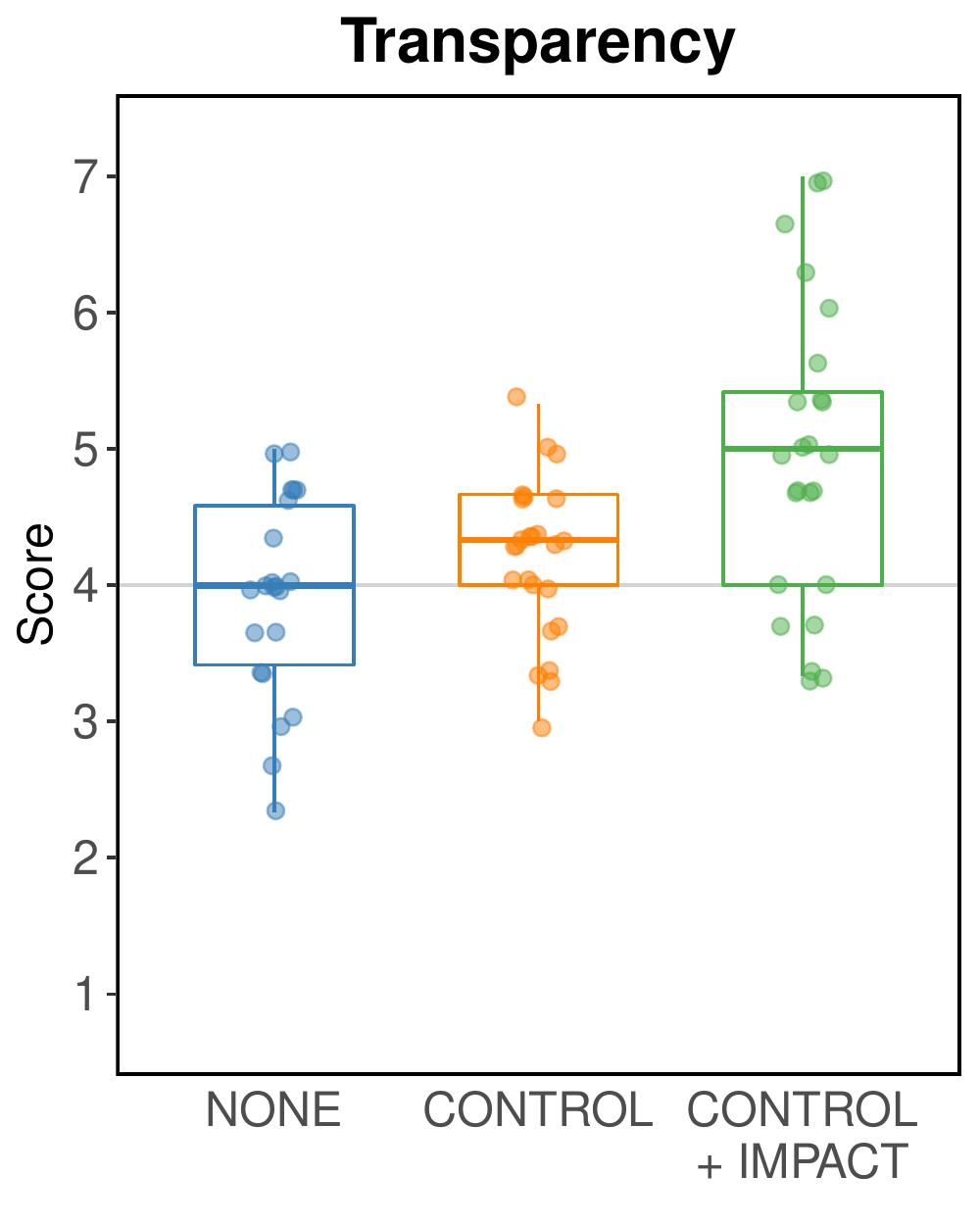}\hfill
        \includegraphics[width=.24\textwidth]{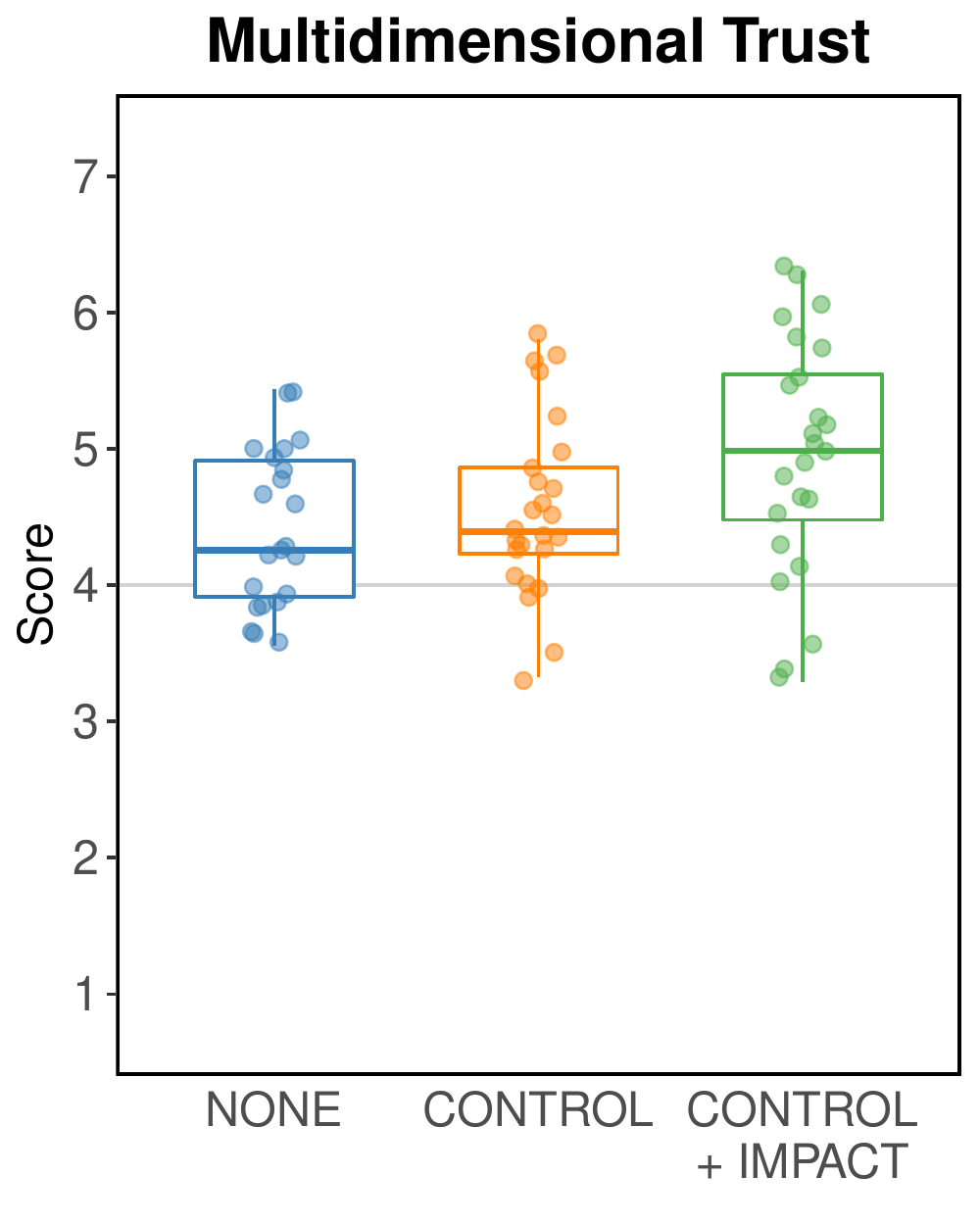}
    \end{minipage}
    \begin{minipage}{.85\linewidth}
        \includegraphics[width=.24\textwidth]{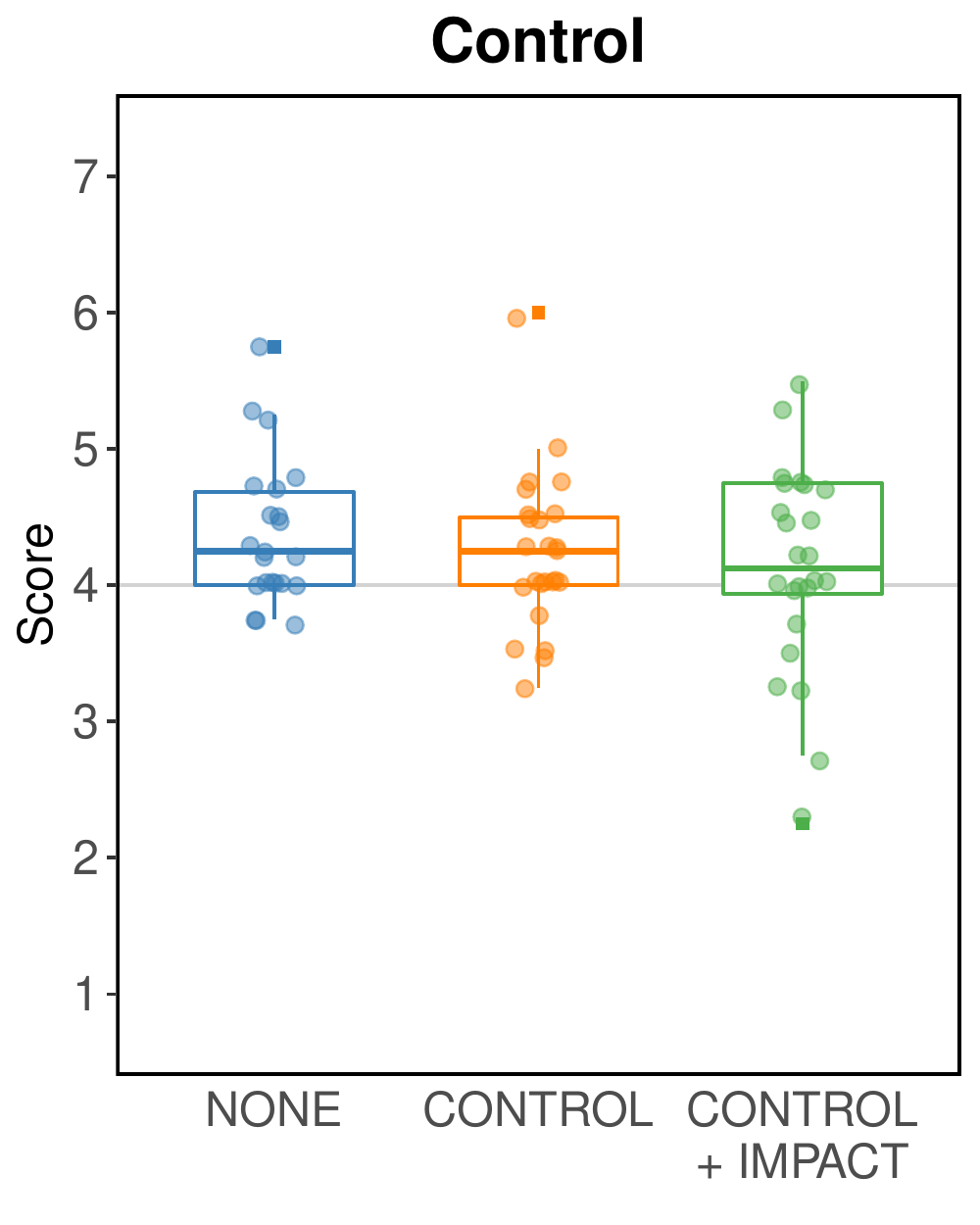}\hfill
        \includegraphics[width=.24\textwidth]{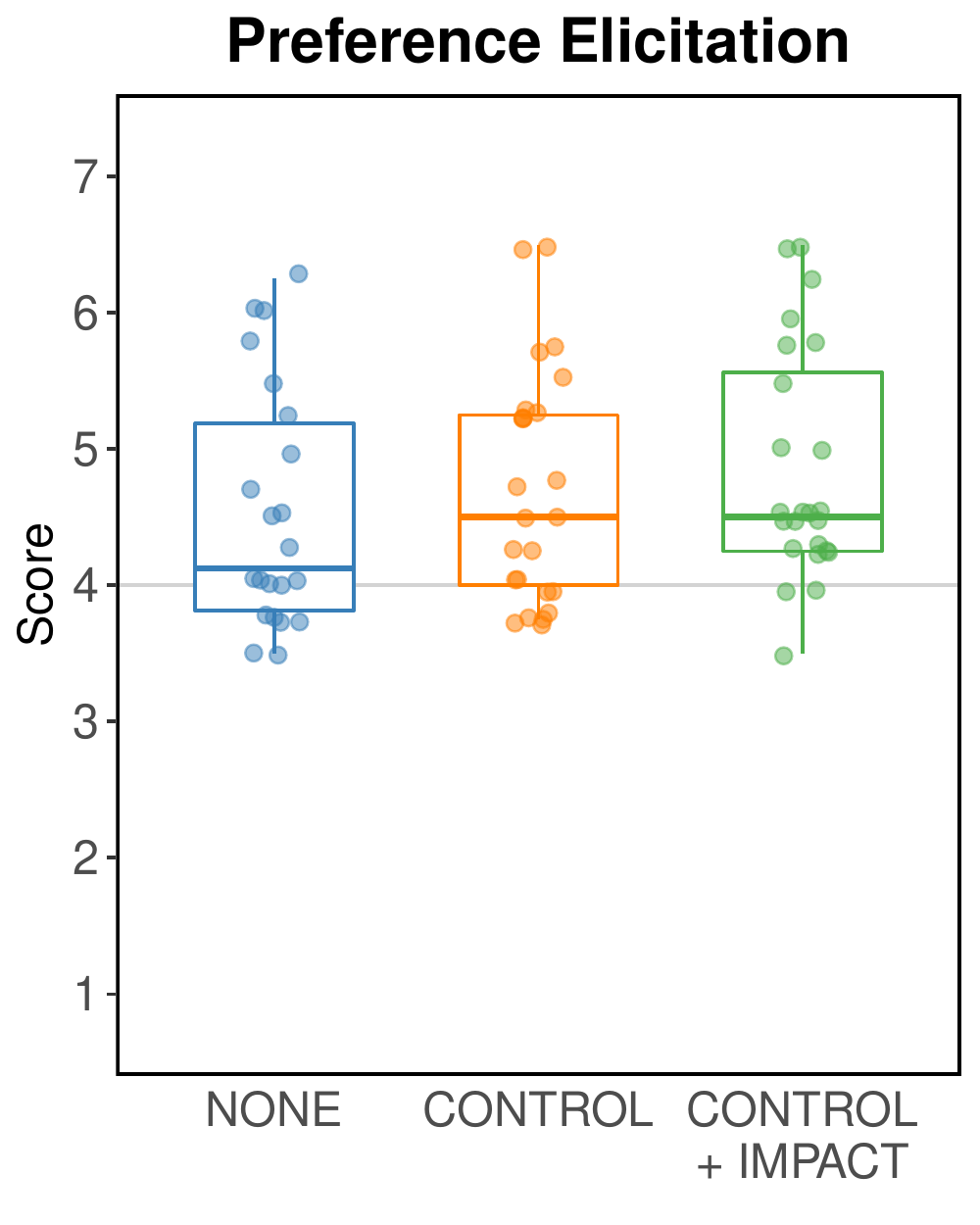}\hfill
        \includegraphics[width=.24\textwidth]{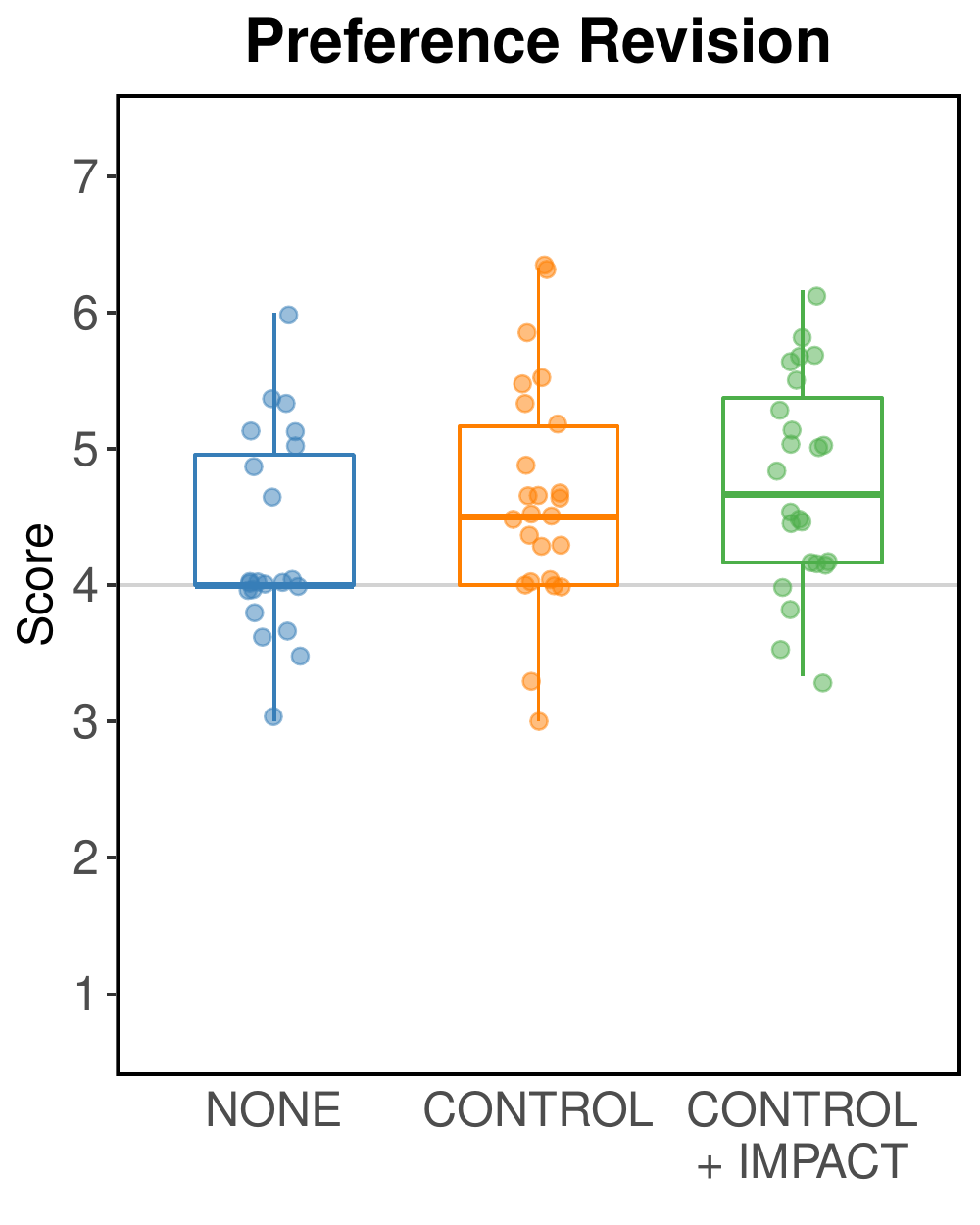}\hfill
        \makebox[.24\textwidth]{}
    \end{minipage}%
    \vspace*{-6pt}
    \caption{Box plots of the responses to the questionnaire in \Cref{tab:questionnaire} for each research group. For visual clarity, the overlaying dot plots are slightly jittered horizontally and vertically.}
    \label{fig:boxplots}
    \Description{Fully described in the caption and text.}
\end{figure*}

\Cref{fig:participants} shows the number of participants per grade, equally distributed over the three research groups. To get a detailed understanding of how participants filled out the Likert-type questions, \Cref{fig:divergingBarchart} depicts the distribution of responses in each research group. In turn, \Cref{fig:boxplots} gives a more aggregated view of participants' responses per research group. Recall that multidimensional trust is the average of trusting beliefs, intention to return, and transparency. Overall, the median scores of all measured constructs lay between neutral and rather agree. ANOVA tests found that competence, integrity, intention to return, control, and preference elicitation did not differ significantly in the three research groups ($p>0.20$).

\subsection{Effects Without Control or Seeing Its Impact}
Qualitative responses confirmed that most participants in \textsc{none} trusted the platform overall. Over one third of the participants seemed to have based their trust on the platform's design and utility: they found that \quote{the website looked professional,} was \quote{good for practising for tests,} was \quote{a good way to practise maths to improve,} and seemed to contain exercises that \quote{fit well to the subject matter.} Furthermore, two participants believed the platform was developed by teachers or experts. Another third of the participants commented on whether exercises had a suitable difficulty level. In case they found exercises well-tailored, participants appeared trusting, for example, \quote{The website looks [...] trustworthy. I also have the feeling that the exercises are of a good level.} Conversely, a few participants appeared distrusting or hesitant because they \quote{often got the same questions they had already answered correctly before,} which gave them the feeling their mastery level stagnated and they could memorise answers. Finally, four participants alluded to potentially different trust perceptions in the long term: \quote{I have not been able to practise and use the site enough, so I cannot give a good final assessment either (at the moment).}

Thirteen participants in \textsc{none} commented on obtaining control over recommended exercises. Apart from one indifferent individual, all of them were in favour of extra control. Only three, however, clarified why: \quote{This allows you to give a bit of direction to what exercises you want yourself. Also, if you perform a bit less well, you still get some more difficult exercises to see what they entail.}

\subsection{Effects of Controlling Recommendations}
The first column in \Cref{table:tests} shows that one-sided tests did not reveal statistical differences between \textsc{none} and \textsc{control} ($p<0.05$). Thus, our sample did not provide evidence against equal means for any measured construct. Only transparency and preference revision were borderline non-significant.

The qualitative responses on trust showed that two thirds of the participants in \textsc{control} seemed trusting and mostly supported that perception by the platform's ability to tailor exercises: \quote{It seems reliable at first sight and it also asks good questions adapted to your maths level}; \quote{It can assess your level and provide further exercises to raise your level}; and \quote{[I trust Wiski] if you can enter your own level.} Furthermore, similar to the responses in \textsc{none}, some participants referred to the platform's \quote{professional} design and utility to \quote{learn something new.} In addition, two participants mentioned repeatedly occurring exercises but did not seem troubled by that: \quote{Wiski knows when I have some difficulties with exercises and when I don't. That's why difficult exercises are recommended again.}

There were, however, also mixed trusting sentiments: while six participants did see benefits in our platform for casual practice, they hesitated to blindly adopt it in the long term for two reasons. First, some were bothered by the algorithmic nature of recommendations: \quote{It's a programme and not a teacher so I don't quite trust it} and \quote{[It's] just an AI [...]. Wiski is good but I'd rather seek advice from a physical person.} Two quotes might explain this sentiment: \quote{It remains a computer system that can always be flawed} and \quote{It only has a limited view of my maths skills.} Second, practice in the context of preparing tests or exams might require the presence of a teacher: \quote{Sometimes teachers have their own way of asking questions and this may not always match the exercises offered by Wiski.}

Furthermore, all respondents in \textsc{control} and \textsc{control+impact} were very positive about the feature to control recommendations. The ability to modify the difficulty level of recommended exercises was especially appreciated to not \quote{get stuck} when \quote{you find the exercises too difficult or too easy} and when \quote{you want to try something harder but also go for something easy once in a while.} Yet, one participant noted that while \quote{the slider is nice to make small adjustments, it's not convenient to specifically choose a new level because [they] wanted to go up 1 level in difficulty and went up 2 levels,} alluding to the five mastery levels depicted in \Cref{fig:initialmastery,fig:impact}. Someone else agreed that it was indeed \quote{difficult to find the perfect level.} Finally, one participant admitted they were \quote{not sure whether [Wiski] understood [they] wanted slightly more difficult exercises} when using the slider.

\begin{figure*}
    \centering
    \includegraphics[width=0.825\linewidth]{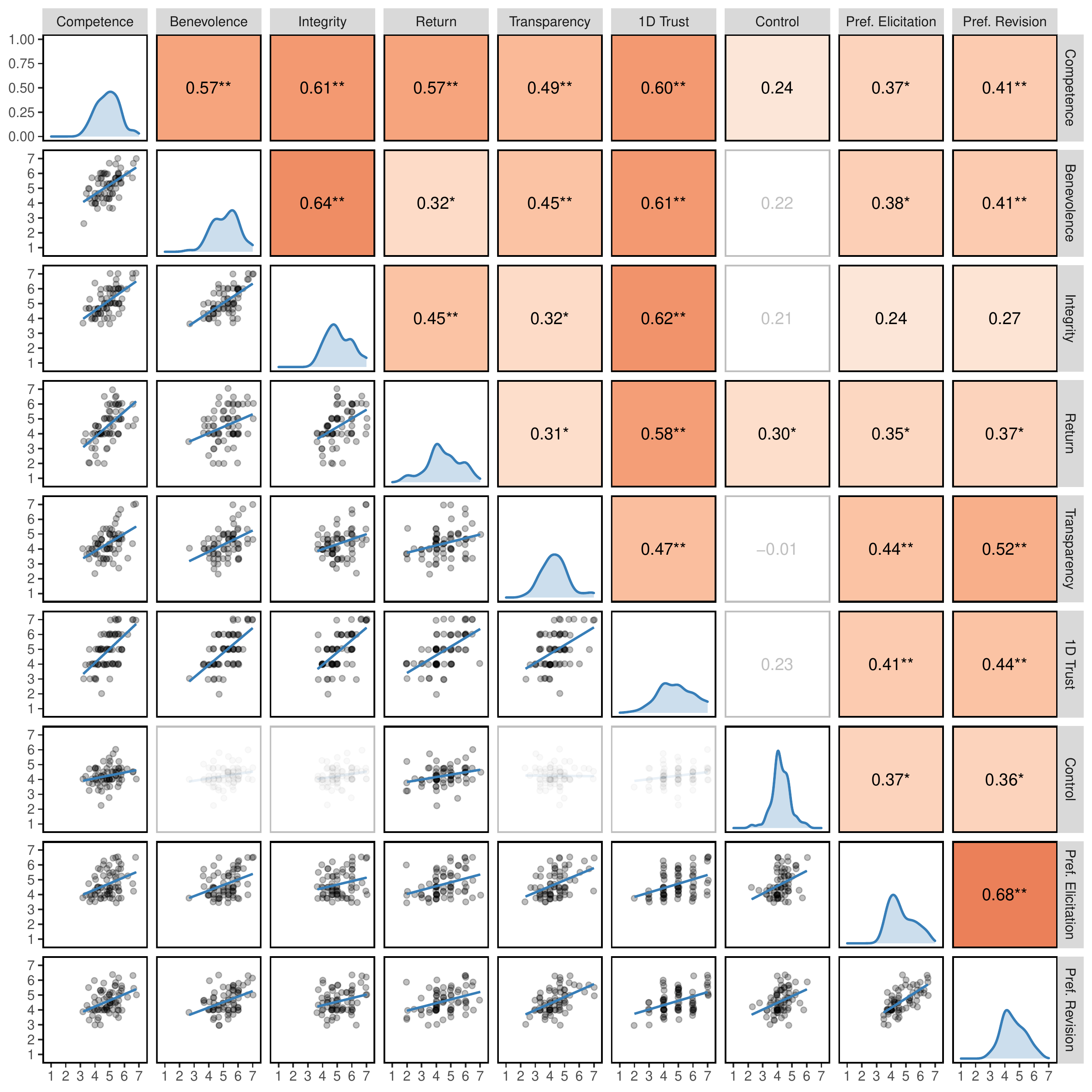}
    \caption{Relations between trust-related and control-related constructs. Lower triangle: dot plots with fitted regression lines. Diagonal: density plot of constructs. Upper triangle: correlations colour-coded by value (*$p<0.01$, **$p<0.001$). Non-significant relations ($p\geq 0.05$) are greyed out.}
    \label{fig:correlations}
\end{figure*}

\subsection{Effects of Visualising the Impact of Control}
The second and third columns in \Cref{table:tests} show the results of comparing \textsc{none} to \textsc{control+impact}, and \textsc{control} to \textsc{control+impact}, respectively. Both one-dimensional trust and multidimensional trust increased significantly $(p < 0.05)$. The latter relates to an increase in two of its components: trusting beliefs and transparency. First, trusting beliefs increased due to higher perceived benevolence. Second, participants perceived the platform as significantly more transparent, with the average score in \textsc{control+impact} lying 1 point higher than in \textsc{none}. Regarding control, however, only preference revision was deemed significantly higher in \textsc{control+impact}, compared to \textsc{none}.

In \textsc{control+impact}, most qualitative responses regarding trust were positive. Similar to \textsc{control}, two thirds of the respondents focused on how well exercises were tailored. Most of these participants trusted the platform and highlighted that exercises were well-tailored: \quote{I think Wiski does give exercises at my level. It's nice that when you get a lot of exercises right, you get more difficult exercises to challenge yourself. You notice that they get harder, therefore I trust the recommended exercises} and \quote{[It's] handy that this platform can estimate your level, the exercises recommended by Wiski are therefore well fit.} Yet, three participants were rather distrustful because exercises seemed ill-tailored or repetitive to them: \quote{I have now made some exercises and have not yet found the level that suits me. So I am more inclined to make exercises in my textbook because I know we should be able to achieve that level.} Other participants seemed to prefer consulting a teacher or using Wiski only for supplementary exercises: \quote{I think Wiski is well-made and does its best to help but I don't think it can really determine my maths level.} Finally, two participants touched upon long-term trust: \quote{It's hard to say whether I fully trust it after just a few exercises.}

Few participants in \textsc{control+impact} commented on the feature to see their control's impact, yet those who did found it useful to see their evolution and current level. In contrast, in \textsc{none} and \textsc{control} together, most participants commented on whether they would have liked a screen similar to \Cref{fig:impact}; all but one would. Many comments tapped into seeing and understanding one's current mastery level: \quote{This can be useful in several ways to see why you are at a certain level} and \quote{Then you can see how well some exercises go.} One participant wrote: \quote{That's pretty handy to see how bad you are at maths.} Another frequent related theme was the possibility to see one's evolution: \quote{That would be useful because then you know how you are progressing.} Finally, one participant brought up motivation, stating \quote{I think this could also be motivating.}

\subsection{Correlations}
\Cref{fig:correlations} shows the relations between all measured trust-related and control-related constructs. Regarding the trust-related constructs, we found that competence, benevolence and integrity were moderately correlated to one another and were equally correlated with one-dimensional trust (all around $r=0.60$). Furthermore, intention to return turned out to be most correlated to competence ($r=0.57$). Regarding the control-related constructs, preference elicitation and preference revision correlated strongly ($r=0.68$), but were barely correlated to control. In fact, control had little to no linear relationship with any of the constructs. Finally, the most correlated pair of trust-related and control-related constructs consisted of transparency and preference revision ($r=0.52$), which is still relatively low as one construct explains only 25\% of the variance in the other.

\aptLtoX[graphic=no,type=html]{\begin{figure}
    \centering
    \includegraphics[width=\linewidth]{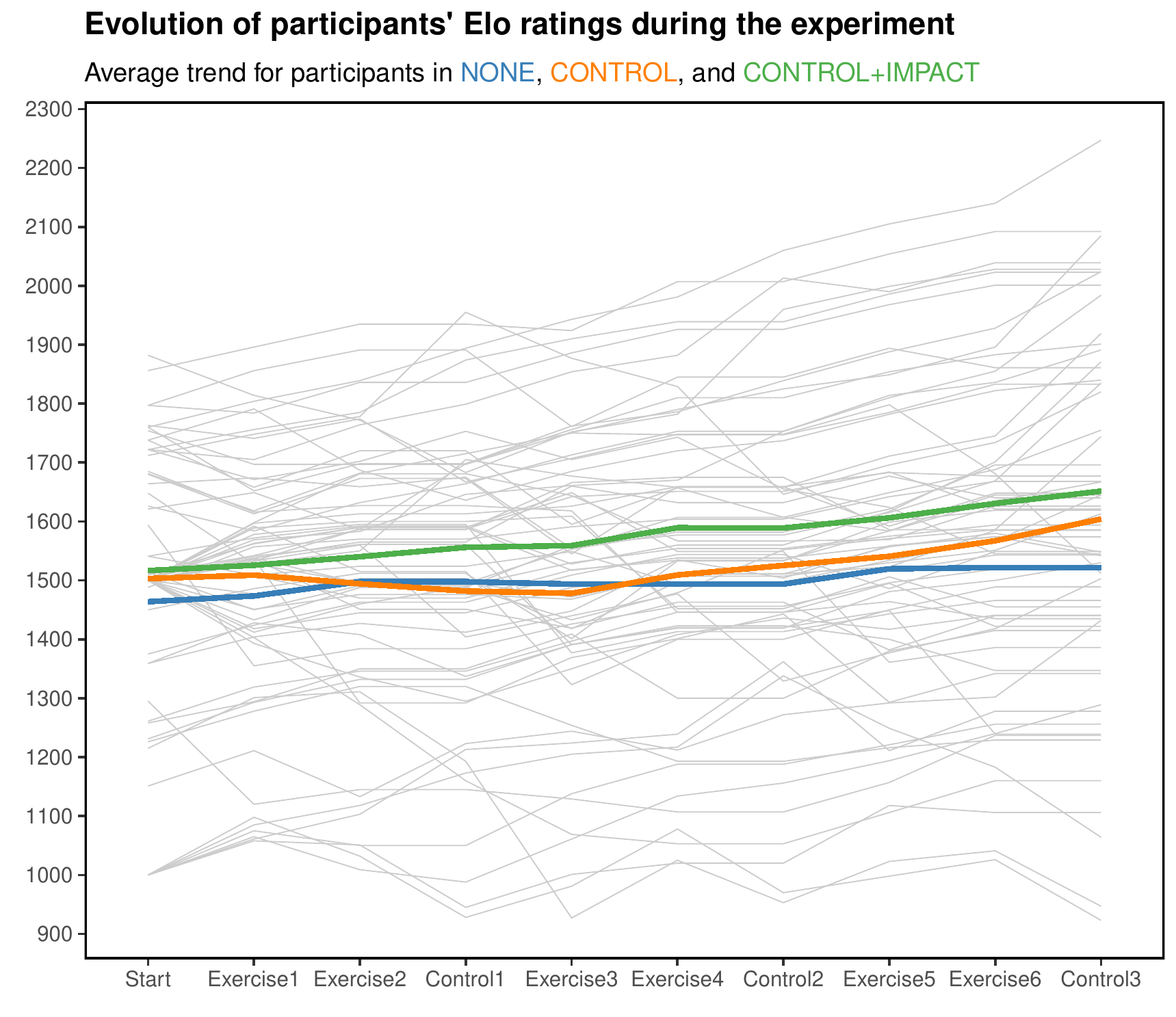}
    \captionof{figure}{Evolution of participants' Elo ratings during the experiment and the average evolution per research group.\vspace*{\baselineskip}}
    \label{fig:evolutionElo}
    \end{figure}
\begin{figure}
    \centering
    \includegraphics[width=\linewidth]{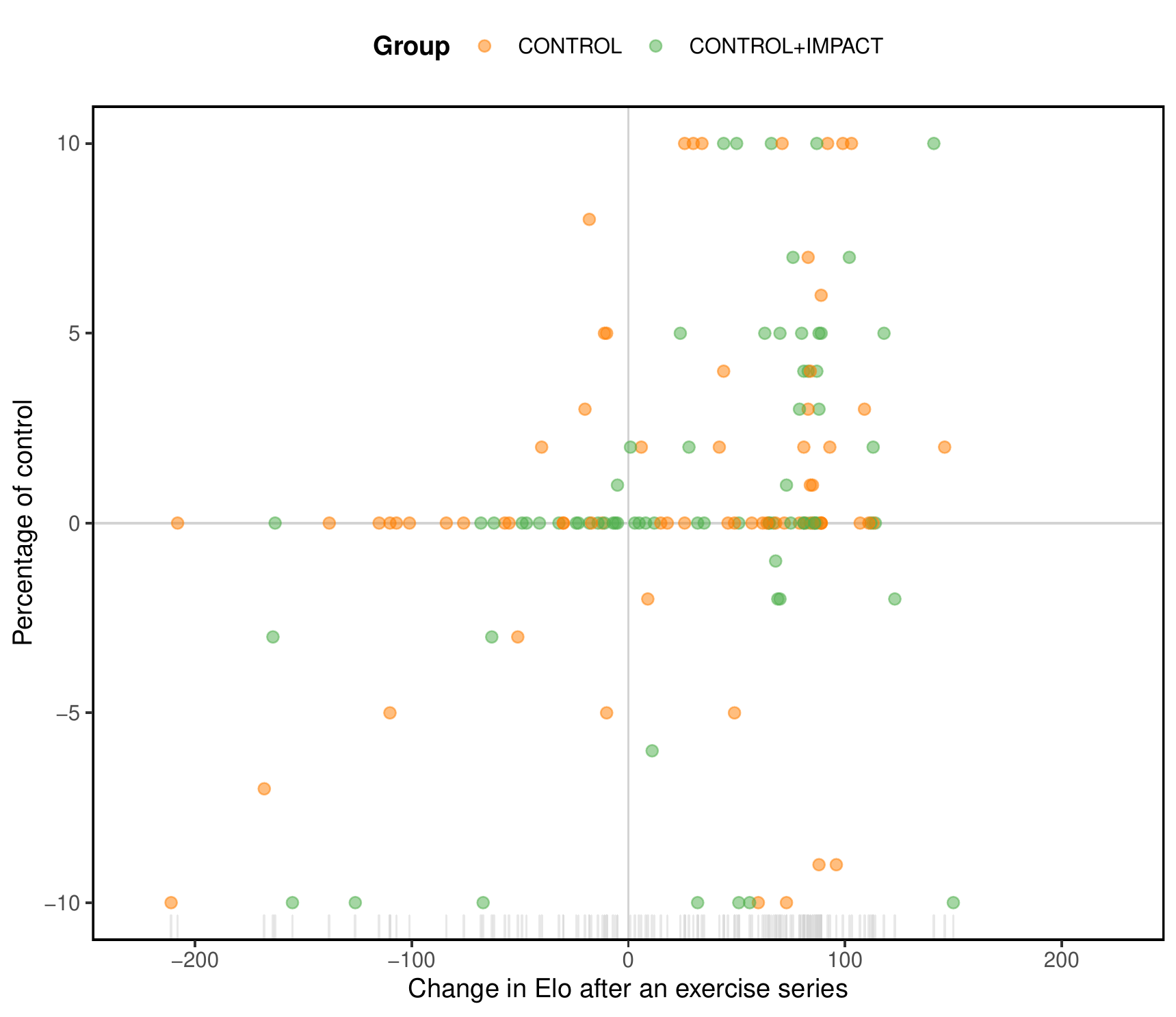}
    \captionof{figure}{Elo changes after an exercise series compared with the control percentage chosen via the slider in \Cref{fig:control} ($\text{Easier}=-10\%$ and $\text{Harder}=10\%$).}
    \label{fig:controlPercentages}
\end{figure}
}{
\begin{figure*}
    \begin{minipage}[b]{.48\linewidth}
    \centering
    \includegraphics[width=\linewidth]{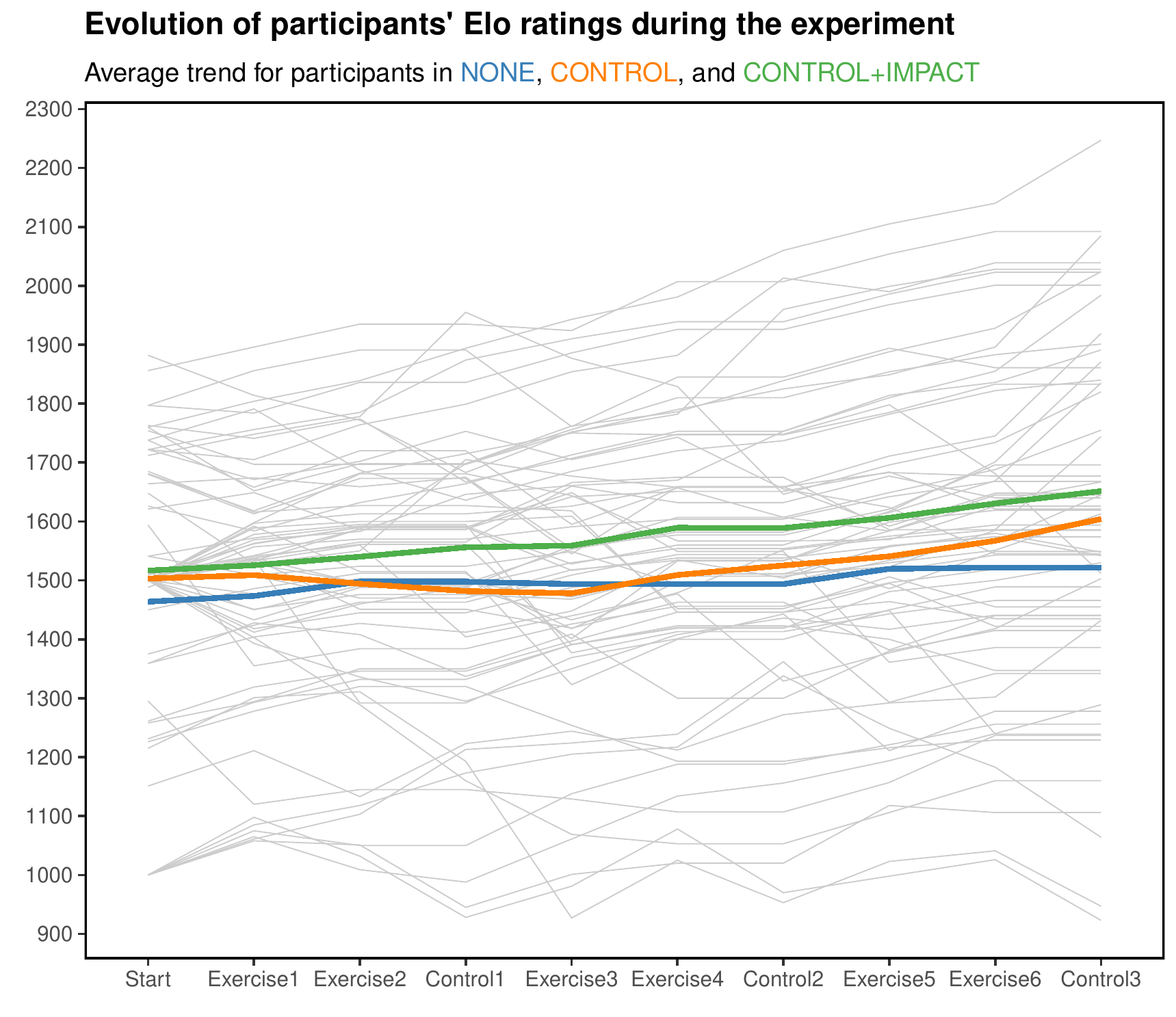}
    \captionof{figure}{Evolution of participants' Elo ratings during the experiment and the average evolution per research group.\vspace*{\baselineskip}}
    \label{fig:evolutionElo}
    \end{minipage}\hfill%
    \begin{minipage}[b]{.48\linewidth}
    \centering
    \includegraphics[width=\linewidth]{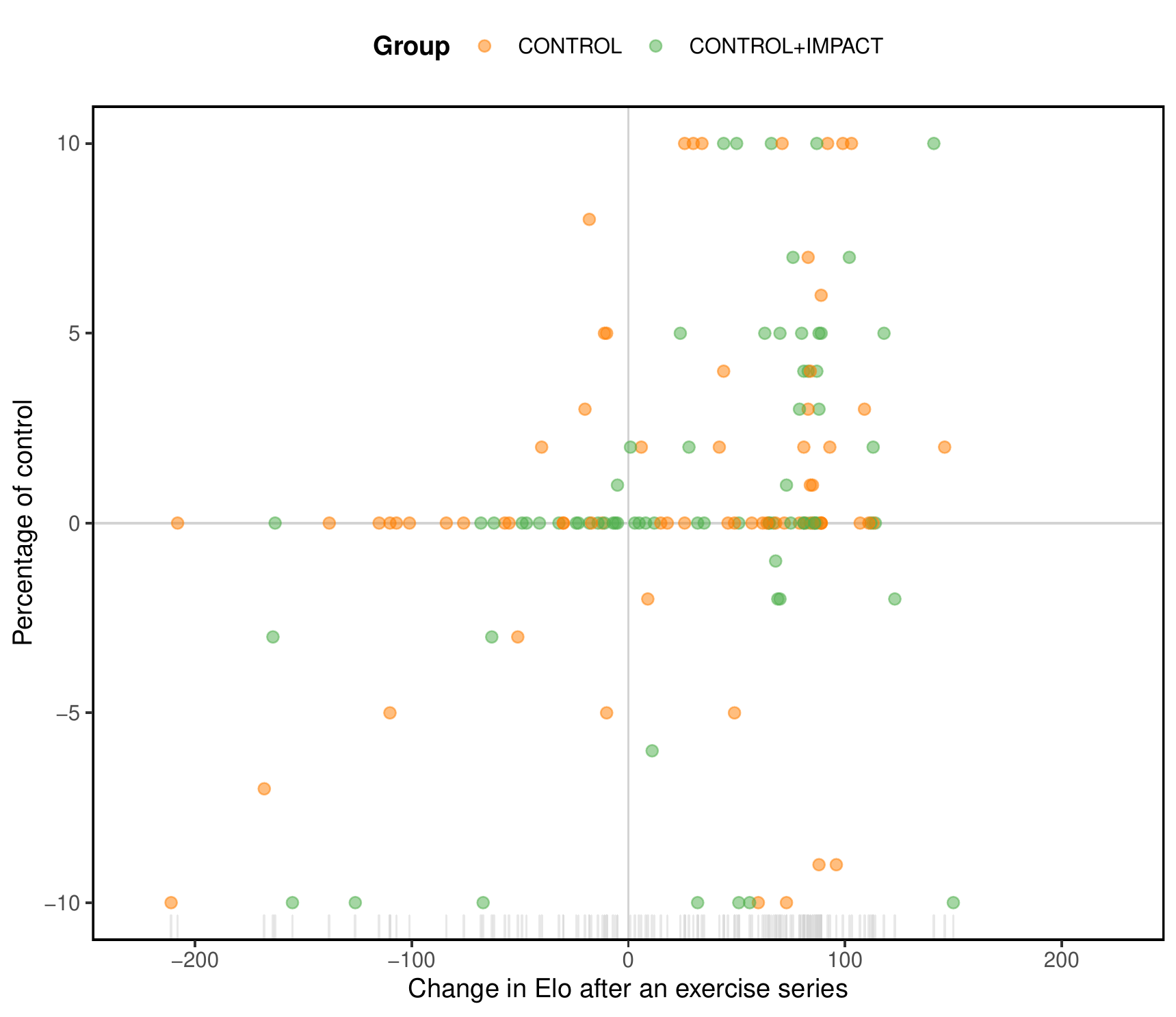}
    \captionof{figure}{Elo changes after an exercise series compared with the control percentage chosen via the slider in \Cref{fig:control} ($\text{Easier}=-10\%$ and $\text{Harder}=10\%$).}
    \label{fig:controlPercentages}
    \end{minipage}
\end{figure*}}

\subsection{Elo Ratings}
\Cref{fig:evolutionElo} shows how participants' Elo ratings evolved during the experiment. In all research groups, the ratings gradually increased and finally participants in \textsc{none} had a lower average increase (58) than participants in \textsc{control} (101) and \textsc{control+impact} (135). Yet, the trends and \Cref{fig:controlPercentages} also show that participants in \textsc{control} and \textsc{control+impact} most often increased their mastery level further after completing an exercise series. Ignoring Elo changes due to control, the average Elo increased with 60 in \textsc{control} and 98 in \textsc{control+impact}. According to one-sided t-tests, however, these average Elo growths were not significantly larger than in \textsc{none} ($p=0.132$). \Cref{fig:controlPercentages} furthermore shows that participants used the control mechanism reasonably: most dots are in the top right quadrant, indicating that participants often further increased their mastery level after a successful exercise series; the left quadrants show that participants rarely boosted their mastery level after an unsuccessful exercise series and instead kept or downgraded it.

\section{Discussion}
This section interprets our results and answers our research questions. Based on our findings, we reflect upon implications for the explainable AI field and real-world e-learning platforms.



\subsection{Sanity Check for Responses About Control}
Before interpreting our results, we take a closer look at the findings regarding control. First, the quantitative results showed no higher sense of control in research groups with control, compared to the baseline without control. This unexpected result could be due to the measurement instrument rather than an actual lack of control: \Cref{fig:divergingBarchart} shows quite polarised responses on Q18--Q21, indicating that the questions may have been interpreted differently because they were too broad. In addition, the qualitative data confirmed that participants in \textsc{control} and \textsc{control+impact} were very aware of the control mechanism and \Cref{fig:controlPercentages} shows that they often used it. Second, preference elicitation was perceived equally amongst the three research groups. This was expected as participants could only indicate initial preferences by setting their initial mastery level and choosing maths topics. Third, also as expected, preference revision increased (almost) significantly when the control mechanism was added, but not when the control's impact was visualised. These observations support the sanity of our results.

\subsection{Control Does Not Affect Trust but Stimulates Self-Reflection}
RQ1 was concerned with how a control mechanism affects trust in our e-learning platform. \Cref{table:tests} contains no evidence for significant effects on any of the measured trust components; only perceived transparency was borderline. The fact that one-dimensional trust did not differ significantly in \textsc{none} and \textsc{control} suggests participants did not consider the control mechanism a major factor for calibrating their trust. 

However, the qualitative responses on trust interestingly revealed more self-reflection. Specifically, while participants in \textsc{none} most often described the platform's utility and design while discussing trust, participants who could control recommendations spontaneously referred twice as much to whether exercises were tailored to their personal mastery level. Some participants even reflected on the recommendation algorithm itself, questioning whether it was as competent as teachers. Thus, our qualitative findings suggest that control mechanisms similar to ours foster awareness of an underlying manipulable algorithm. Growing such awareness seems very valuable in a world that becomes permeated by applications relying on algorithmic decision-making, so future experiments could investigate whether and why this effect holds in larger samples. One plausible explanation could be that controlling mechanisms are uncommon in current e-learning platforms and therefore caught participants' attention.

\subsection{Seeing the Impact of Control Grows Trust}
RQ2 asked how visualising the impact of control influences adolescents' trust in our e-learning platform. Our results showed a significant increase in one-dimensional trust, which suggests that the visualisation played a big role in growing trust. Multidimensional trust also increased significantly, partly due to a higher perceived benevolence. This could be explained by the following observation: \Cref{fig:evolutionElo,fig:controlPercentages} show that most exercise series led to an increase in Elo rating, which implies that participants in \textsc{control+impact} mostly saw increasing mastery evolutions. Thus, it seems plausible that participants who saw the visualisation considered our platform as more benevolent than participants who did not.


\subsection{Visualising the Impact of Control is a Kind of Explanation}
The most heavily changed trusting component was transparency: participants who saw their control's impact visualised considered the recommendations as more transparent. This suggests that participants experienced the visualisation as a sort of explanation. However, at first sight, it is not entirely clear \textit{what} part of the recommendation process this explanation clarified for them. Comparing the responses for Q15--Q17 in \Cref{fig:divergingBarchart}, we observe that roughly half of the responses for Q17 were negative, whereas most responses for Q15 and Q16 were positive. This seems to imply that participants did not view the visualisation of their control's impact as a \textit{direct} explanation for why they received specific recommendations; which they indeed should not have. Rather, the visualisation arguably acted as an \textit{indirect} explanation: participants felt they had a better understanding of why recommendations were suitable for them because they could repeatedly see how the e-learning platform estimated and modified their mastery level. Overall, visualising the impact of control seems to have reinforced participants' mental model~\cite{johnson-laird1983mental,kulesza2012tell} of the recommendation system by gradually clarifying the behaviour of a crucial component of the recommender, namely iterative estimation of learners' mastery level.

\subsection{Implications for Explainable AI Research}
Our findings potentially have interesting research implications for the broader field of explainable AI. First, visual explanations intended for a lay audience may not need to explain complete algorithms in detail. Instead, explaining crucial components could suffice when complemented with a global reasoning rationale of the algorithm. In our case, this global reasoning rationale was provided as a simple sentence on the practice page (\quote{You will automatically get the two exercises that best suit your level} in \Cref{fig:explanations}). Another provoking idea is that control by itself could increase transparency. This turned out to be the case in our sample, although the increase was borderline non-significant. We hypothesise that on our platform the combination of exercising control and seeing the difficulty level of subsequently recommended exercises acted as a kind of \textit{model inspection}~\cite{guidotti2019survey}. In other words, participants could steer the recommendation algorithm and then see the impact on outcomes of the recommendation algorithm. If future research could confirm our hypothesis, this might be one of the earliest examples of effective model inspection with adolescents. Third, the qualitative responses regarding motivation open up research tracks on whether (visual) explanations can inherently motivate students to, for example, practise more, challenge themselves more, or -- being hopeful -- even appreciate maths more as a whole.

\subsection{Taking a Step Back: Technology-Enhanced Learning and Control}
Before we conclude, we briefly reflect upon control in e-learning. How much control should students get and does that imply taking control away from teachers? Should students always see their control's impact with respect to their mastery level?

Overall, students received our platform's control mechanism enthusiastically and seemed to have used it reasonably. The faster increase in Elo rating for students with control also suggests that the control mechanism allowed them to more quickly converge towards exercises with difficulty levels that best suited them. Moreover, the control mechanism and its accompanying impact visualisation seemed to have prompted students to think more consciously about which difficulty levels they could handle and how their mastery level changed. This is an important metacognitive skill, which is crucial in self-regulated learning~\cite{zimmerman1990selfregulated}. For these reasons, we believe giving control to students can be an asset for e-learning platforms.

Yet, we see at least two nuances. First, giving too much control to students can be disadvantageous when it causes discomfort because of the responsibility it entails~\cite{mabbott2006student}. In addition, students could abuse control over their mastery level in evaluative contexts: artificially decreasing their mastery level could allow them to obtain higher success rates when solving exercises, and artificially increasing it could trick inattentive teachers into overestimating their abilities. Thus, it is important to balance the amount of control with factors such as pedagogical responsibility and the use context and to not overly rely on Elo ratings for evaluation purposes. Second, providing students with control does not make teachers redundant. In our study, participants highlighted the still valuable role of teachers: providing extra feedback on students' progress, and verifying that exercises on the e-learning system are aligned with the curriculum and their usual style of interrogating. Furthermore, by monitoring or adapting students' mastery levels, teachers could additionally guide students who under- or overestimate themselves because of the Dunning-Kruger effect \cite{dunning2011chapter}.

Our visualisation of how control affected mastery level, and thus recommended exercises, was well-received too. However, some comments regarding motivation made us realise the potentially demotivating effects of frequently showing downward evolutions in students' mastery levels. Therefore, we argue that visualisations related to mastery level should be shown sufficiently infrequent to avoid potential negative motivational effects, yet frequently enough to allow intervention in case of learning issues. Such interventions could be facilitated by e-learning platforms in the form of alerts that inform students when it seems advisable they ask teachers for additional support. In line with teachers' desires in our pilot study, those alerts could also be shown to teachers so they can intervene, similar to existing work in the learning analytics community~\cite{denden2019imoodle,akcapinar2019using}.

\subsection{Limitations and Future Work}
Our research has several limitations which restrict how well our findings generalise. First, our sample was relatively small so some findings may not hold in larger studies and we could not investigate differences between age groups. Although we controlled for multiple testing by only conducting t-tests when ANOVA indicated a group-wise difference, false positive differences could remain. Second, since our study was not focused on developing a highly accurate recommender system, we generated recommendations with a simple Elo rating system. More sophisticated algorithms such as multivariate Elo-based models~\cite{abdi2019multivariate} or knowledge tracing~\cite{guo2021enhancing} could be considered for platforms deployed in the real world, especially because competence is rather highly correlated to intention to return (see \Cref{fig:correlations}). Third, the mechanism to steer recommendations was quite simple and only affected recommended exercises indirectly by altering mastery levels. Future studies with adolescents in e-learning could further study more advanced control mechanisms that affect recommendations directly, for example steering through interactive visualisations. Fourth, as our study was conducted in a class context, it is possible that some students noticed that their peers were shown a different version of our platform. Although we did not observe copying during the study, we are wary of adolescents' resourcefulness to copy and the bias it may have entailed. Fifth, as some participants indicated, our study was restricted to capturing trust while participants were arguably still familiarising themselves with the recommender and control mechanism. In this \textit{learning phase}~\cite{yu2017user}, trust perceptions can change briskly, for example due to encountering unexpected behaviour such as repeated recommendations~\cite{nourani2020role,yu2017user,holliday2016user}. Thus, as briefly using our platform might have hampered reliable long-term trust assessment, our results should be interpreted cautiously. Sixth, our results regarding transparency relied on self-reported understanding. Future research could complement transparency measurements with testing effective understanding, for example through adjusted tasks. Overall, we hope our suggestions help to pursue research into providing adolescents with control over recommendations in e-learning.

\section{Conclusion}
Our research explored how a control mechanism for steering recommended exercises and a visualisation of the control's impact influence adolescents' trust in an e-learning platform. We measured trust both with a single Likert-type question and as a multidimensional construct of trusting beliefs, intention to return, and perceived transparency. In addition, we collected qualitative feedback to further contextualise students' responses. Our randomised controlled experiment with 76~middle and high school students showed that our control mechanism did not significantly change any trusting perception. However, adolescents appreciated the feature and seemed to reflect more upon their mastery level and the recommendation system, which is highly favourable in the context of self-regulated learning. Furthermore, visualising the control's impact did increase trust and perceived understanding, which suggests several implications for the broader field of explainable AI. In sum, even though our study had limitations, we hope our methods, designs, and findings inspire other researchers to further explore the link between control mechanisms, explainable AI, and motivational techniques, especially in e-learning and targeting adolescents.

\begin{acks}
We thank all students who participated in our study, their parents for consenting, and their teachers for inviting us into their classrooms. This work was supported by Research Foundation–Flanders (FWO, grant G0A3319N), Flanders Innovation \& Entrepreneurship (imec.icon grant HB.2020.2373), and KU Leuven (grant C14/21/072).
\end{acks}

\bibliographystyle{ACM-Reference-Format}
\bibliography{IUI2023}

\newpage
\onecolumn
\appendix
\section{Post-Study Questionnaire}

\begin{table}[h]
	\caption{The questionnaire that participants filled out at the end of the study. All questions were evaluated on a 7-point range, and questions Q19, Q20, and Q25 were reverse-scored. The italic group names are for reference; participants did not see them.}
	\label{tab:questionnaire}
	\footnotesize
	\begin{tabularx}{\textwidth}{@{}lXX@{}}
		\toprule
		\textbf{No.} & \textbf{English version}                                                                   & \textbf{Dutch version}                                                                              \\
		\midrule
		\multicolumn{3}{@{}l}{\textit{Competence}}                                                                                                                                                                           \\
		Q1           & Wiski is like an expert (for example, a teacher) for recommending maths exercises.           & \dutch{Wiski is zoals een expert (bv. een leerkracht) in wiskunde-oefeningen aanraden.}                 \\

		Q2           & Wiski has the expertise (knowledge) to estimate my maths level.                              & \dutch{Wiski heeft de expertise (kennis) om mijn wiskundeniveau te kunnen inschatten.}                  \\

		Q3           & Wiski can estimate my maths level.                                                           & \dutch{Wiski kan mijn wiskundeniveau inschatten.}                                                       \\

		Q4           & Wiski understands the difficulty level of maths exercises well.                              & \dutch{Wiski begrijpt de moeilijkheidsgraad van wiskunde-oefeningen goed.}                              \\

		Q5           & Wiski takes my maths level into account when recommending exercises.                         & \dutch{Wiski houdt rekening met mijn wiskundeniveau om oefeningen aan te raden.}                        \\
		\midrule
		\multicolumn{3}{@{}l}{\textit{Benevolence}}                                                                                                                                                                          \\
		Q6           & Wiski prioritises that I improve in maths.                                                   & \dutch{Wiski zet op de eerste plaats dat ik vorderingen maak in wiskunde.}                              \\

		Q7           & Wiski recommends exercises so that I improve in maths.                                       & \dutch{Wanneer Wiski oefeningen aanraadt, doet Wiski dat zodat ik vorderingen maak in wiskunde.}        \\

		Q8           & Wiski wants to estimate my maths level well.                                                 & \dutch{Wiski wil mijn wiskundeniveau goed inschatten.}                                                  \\
		\midrule
		\multicolumn{3}{@{}l}{\textit{Integrity}}                                                                                                                                                                            \\
		Q9           & Wiski recommends exercises as correctly as possible.                                        & \dutch{Wiski raadt oefeningen op een zo correct mogelijke manier aan.}                                  \\

		Q10          & Wiski is honest.                                                                            & \dutch{Wiski is eerlijk.}                                                                               \\

		Q11          & Wiski makes integrous recommendations.                                                      & \dutch{Wiski maakt oprechte aanbevelingen.}                                                             \\
		\midrule
		\multicolumn{3}{@{}l}{\textit{Trust (one-dimensional)}}                                                                                                                                                              \\
		Q12          & I trust Wiski to recommend me maths exercises.                                               & \dutch{Ik vertrouw Wiski om mij wiskunde-oefeningen aan te raden.}                                      \\
		\midrule
		\multicolumn{3}{@{}l}{\textit{Intention to return}}                                                                                                                                                                  \\
		Q13          & If I want to solve maths exercises again, I will choose Wiski.                               & \dutch{Als ik nog eens online wiskunde-oefeningen maak, dan kies ik voor Wiski.}                        \\
		Q14          & If I want to be recommended maths exercises again, I will choose Wiski.                      & \dutch{Als ik nog eens wiskunde-oefeningen aangeraden wil krijgen, dan kies ik voor Wiski.}             \\
		\midrule
		\multicolumn{3}{@{}l}{\textit{Transparency}}                                                                                                                                                                         \\
		Q15          & I understood why the exercises were recommended to me.                                          & Ik begreep waarom de oefeningen aan mij werden aanbevolen.                                              \\
		Q16          & Wiski helps me understand why the exercises were recommended to me.                        & Wiski helpt mij te begrijpen waarom de oefeningen aan mij werden aanbevolen.                            \\
		Q17          & Wiski explains why the exercises are recommended to me.                            & Wiski legt uit waarom de oefeningen aan mij worden aanbevolen.                                          \\
		\midrule
		\multicolumn{3}{@{}l}{\textit{Control}}                                                                                                                                                                              \\
		Q18          & I feel in control of telling Wiski what I want.                                   & Ik heb het gevoel dat ik Wiski kan vertellen wat ik wil.                                                \\
		Q19          & I don’t feel in control of telling Wiski what I want.                                  & Ik heb niet het gevoel dat ik Wiski kan vertellen wat ik wil.                                           \\
		Q20          & I don’t feel in control of specifying and changing my preferences.                          & Ik heb niet het gevoel dat ik controle heb over het omschrijven en veranderen van mijn voorkeuren.      \\
		Q21          & Wiski seems to control my decision process rather than me.                             & Wiski lijkt mijn keuzeproces te controleren in plaats van ikzelf.                                       \\
		\midrule
		\multicolumn{3}{@{}l}{\textit{Preference elicitation}}                                                                                                                                                               \\
		Q22          & Wiski provides an adequate way for me to express my preferences.                  & Wiski laat me op een geschikte manier mijn voorkeuren aangeven.                                         \\
		Q23          & I found it easy to tell Wiski about my preferences.                                    & Ik vond het gemakkelijk om Wiski over mijn voorkeuren te vertellen.                                     \\
		Q24          & It is easy to learn to tell Wiski what I like.                                         & Het is gemakkelijk om te leren hoe ik Wiski kan vertellen wat ik leuk vind.                             \\
		Q25          & It required too much effort to tell Wiski what I like.                                 & Het kostte te veel moeite om Wiski te vertellen wat ik leuk vind.                                       \\
		\midrule
		\multicolumn{3}{@{}l}{\textit{Preference revision}}                                                                                                                                                                  \\
		Q26          & Wiski provides an adequate way for me to revise my preferences.                   & Wiski laat me op een geschikte manier mijn voorkeuren aanpassen.                                        \\
		Q27          & I found it easy to make Wiski recommend different things to me.                        & Ik vond het gemakkelijk om Wiski mij verschillende dingen te laten aanbevelen.                          \\
		Q28          & It is easy to train Wiski to update my preferences.                                    & Het is gemakkelijk om Wiski te leren mijn voorkeuren aan te passen.                                     \\
		Q29          & I found it easy to alter the recommended exercises due to my preference changes. & Ik vond het gemakkelijk om de aanbevolen oefeningen te wijzigen met mijn voorkeursveranderingen.        \\
		Q30          & It is easy for me to inform Wiski if I dislike/like recommended exercises.              & Het is voor mij gemakkelijk om Wiski te laten weten of ik de aanbevolen oefeningen leuk/niet leuk vind. \\
		Q31          & It is easy for me to get a new set of recommended exercises.                                      & Het is voor mij gemakkelijk om een nieuwe reeks aanbevolen oefeningen te krijgen.                       \\
		\bottomrule
	\end{tabularx}
\end{table}

\end{document}